\documentclass[%
 reprint,
 amsmath,amssymb,
 aps,
]{revtex4-1}

\usepackage{graphicx}
\usepackage{dcolumn}
\usepackage{bm}

\usepackage[version=3]{mhchem} 

\begin{document}

\title{Temporal pattern recognition through analog molecular computation}

\author{Jackson O'Brien}
\author{Arvind Murugan}\email{amurugan@uchicago.edu}
\affiliation{The James Franck Institute and Department of Physics,\\  University of Chicago, Chicago, IL 60637}

\begin{abstract}
Living cells communicate information about physiological conditions by producing signaling molecules in a specific timed manner. Different conditions can result in the same total amount of a signaling molecule, differing only in the pattern of the molecular concentration over time. Such temporally coded information can be completely invisible to even state-of-the-art molecular sensors with high chemical specificity that respond only to the total amount of the signaling molecule.  Here, we demonstrate design principles for circuits with temporal specificity, that is, molecular circuits that respond to specific temporal patterns in a molecular concentration. We consider pulsatile patterns in a molecular concentration characterized by three fundamental temporal features - time period, duty fraction and number of pulses. We develop circuits that respond to each one of these features while being insensitive to the others. We demonstrate our design principles using abstract Chemical Reaction Networks and with explicit simulations of DNA strand displacement reactions. In this way, our work develops building blocks for temporal pattern recognition through molecular computation.
\end{abstract}

\maketitle
Recent breakthroughs in synthetic biology have led to molecular sensors that  report on the local environment in cells by detecting signaling molecules with high \textit{chemical} specificity \cite{Song2008-tv, Modi2009-li,Spichiger-Keller2008-qr,Chen2013-ry}.  

However, by themselves, such sensors can be completely blind to temporally coded information in cells and tissues. In fact, living cells often communicate information about physiological conditions by producing a signaling molecule in a specific timed manner \cite{Purvis2013-io,Hao2011-fr,Hansen2013-cs}.  For example, many rapid pulses of nuclear p53 in mammalian cells indicates $\gamma$ radiation damage and leads to cell cycle arrest, while a longer sustained single pulse of nuclear p53 indicates UV damage and leads to programmed cell death \cite{Purvis2013-io}. Thus different biological conditions can result in the same total amount of a signaling molecule, differing only in the pattern of the molecule's concentration over time \cite{Hansen2016-fo}. Such biological conditions cannot be distinguished by a sensor that responds to the total amount (or exposure) to a target molecule, even if the sensor has high chemical specificity. 

\begin{figure*}
    \centering
    \includegraphics{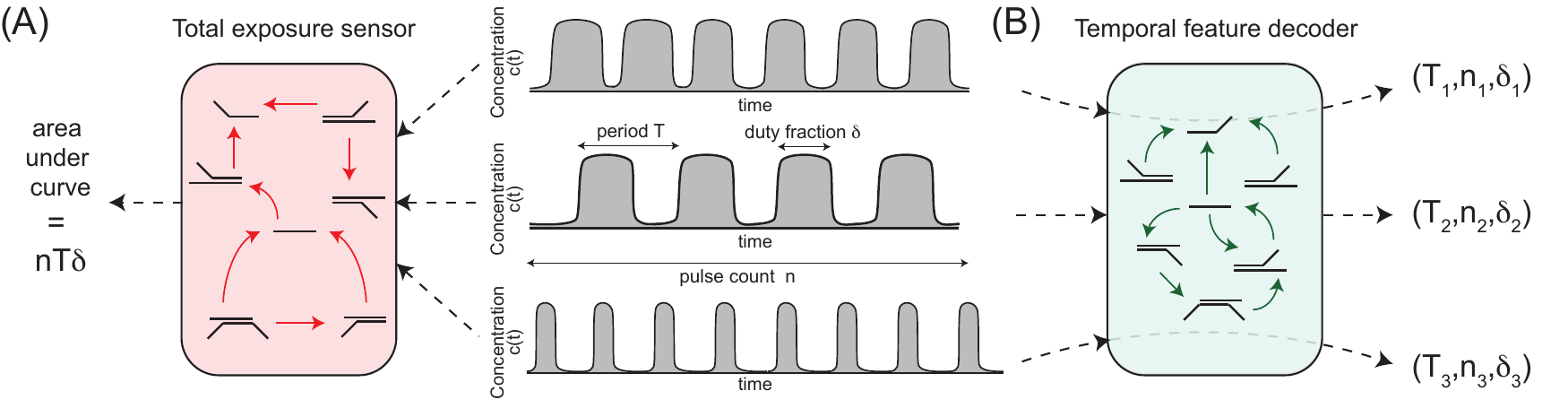}
    \caption{Schematic of temporal feature extraction. The three time-varying patterns of a molecular concentration $ c(t)$ differ in the number of pulses $n$, the time period $T$ (i.e., time between onset of pulses), and duty fraction $\delta$ (i.e., pulse width relative to time period) but have the same total integrated exposure $n T \delta $. (a) Circuits that respond to the total integrated exposure are thus unable to distinguish the three signals. (b) We seek chemical reaction networks that can extract independent temporal features $n,T,\delta$ and report each of them quantitatively. For example, such a sensor must be able to count pulses  without regard to pulse width and report duty fraction without regard to the time period.}
    \label{fig:schematic}
\end{figure*}


In this article, we demonstrate design principles for molecular circuits with \emph{temporal} specificity, i.e., molecular circuits that respond to specific temporal features in the concentration of an input molecule, instead of the total exposure to that input molecule. We show chemical reaction networks satisfying these principles and derive constraints on their rate constants. We then find explicit implementations of these abstract reaction networks using DNA strand displacement reactions and verify the performance of such DNA circuits using simulation software explicitly designed for this purpose \cite{Lakin2011-lb}. We anticipate that this work will also be of use for other synthetic biology constructs based on enzymes and transcriptional gates \cite{Del_Grosso2015-ce, Kim2011-bz} and for analyzing naturally occurring temporal decoding mechanisms in cells \cite{Purvis2013-io}.

While time-varying signals are high dimensional and can vary in endless ways, we restrict our study here to pulsatile patterns as shown in Figure \ref{fig:schematic}. We focus on three fundamental temporal features that characterize such signals - time period $T$ (i.e., time between onset of pulses), duty fraction $\delta$ (i.e., width of each pulse relative to duration between pulses) and the number of pulses $n$ (i.e. length of the pulse train). We construct circuits that respond to each one of these three independent features while being insensitive to other features. 

\textbf{Analog computing with transients}

Recognizing temporal patterns can be interpreted as an analog operation that is naturally suited for molecular computation.
While molecular circuits of digital gates have solved remarkable problems\cite{Qian2009-lq,Qian2011-yd, Qian2011-jh} and even mimicked deep neural networks\cite{Cherry2018-qf},
several recent papers \cite{Daniel2013-sn,Song2018-ba, Song2016-ex, Fern2017-cd,Yordanov2014-tn, Agrawal2015-bt,Genot2013-rd,Virinchi2017-pv, Virinchi2018-bw,ZadorinEstevez2015} 
have shown that analog computations are naturally suited for molecular circuits. Analog operations represent information directly in continuous physical variables like concentration or time intervals\cite{Sarpeshkar2014-qx,Sarpeshkar1998-yx}. Such direct representations can make analog devices smaller and simpler than corresponding digital computation with binary encoding.  The temporal pattern recognition problem studied here is naturally suited to analog computation since the input itself is temporally encoded\cite{Song2018-ba, Song2016-ex}. Indeed, some of the earliest applications of analog computing were to similar temporal decoding problems in electrical circuits, such as AM or FM decoding in radios \cite{Sarpeshkar1998-yx}. While there are theoretical limitations on chemical timers \cite{Doty2013}, several recent works on DNA-based and other molecular computing have addressed related temporal questions. Even a simple boolean AND-gate can be considered a coincidence detector between two signals\cite{Seelig2006-tp} and a temporal sequence detector was recently demonstrated \cite{Kishi2018-qq}, though these constructs do not detect specific temporal intervals. An AND gate and a time-delay mechanism can be combined in a Coherent Feed-Forward Loop that serves as a pulse width discriminator \cite{Alon2007-qt}. Recent work has shown how to convert oscillatory signals into other stereotyped temporal patterns \cite{Agrawal2015-bt,Cuba_Samaniego2018-ja}.

As in these recent analog circuits, our approach here seeks to turn a weakness inherent to molecular devices in the context of digital computation, into a strength for analog computation. Digital computation through molecules requires all internal transients to rapidly die out, so the system can be approximately described by discrete steady state concentrations. In contrast, even on a digital computer, temporal pattern recognition problems are best solved using recurrent networks with internal transients\cite{Graves2013-qs}, unlike, say, feed-forward networks used for image recognition. Such finite timescale transients serve as natural ‘rulers’ to measure and process information encoded as time intervals. Thus our work exploits the inevitable transients in real molecular systems to perform useful computation. 

Natural temporal patterns can be more complex than the family studied here and some applications might requiring detecting more complex features or combinations of features presented here. Such complex temporal processing can be performed by combining the basic motifs and design principles introduced here. In this way, our work lays the foundation for temporal pattern recognition through analog molecular computation.

\section{Results and discussion}

We consider time varying patterns $c(t)$ in the concentration of an input molecule $I$ composed of pulses like those shown in Fig.\ref{fig:schematic}. The patterns shown can be described by three independent numbers; the number of pulses ($n$), the time period $T$, and the duty fraction $\delta \in [0,1]$ (i.e., pulse width relative to time period $T$). 

As shown in Fig.\ref{fig:schematic}, the same total amount of input molecule $I$ can be spread out over time in many different ways. The different patterns shown have distinct $n,T,\delta$ but have the same total area $=n T \delta$. Hence a naive sensor that is only sensitive to the total amount of $I$ would respond in the same way to all patterns shown. 

Here, we seek decoders that can report on each of these temporal features independently - e.g., the duty fraction readout should be independent of the time period and the number of pulses while the number of pulses readout should not depend on the width or separation of the pulses. In what follows, we demonstrate the decoder for each of these temporal features - $n,T,\delta$ - one at a time. In each case, we first demonstrate a high-level Chemical Reaction Network of the architecture and then show the DNA strand displacement-based implementations. The three independent numbers $n,T,\delta$ completely characterize the family of patterns shown here, i.e., they form a complete independent `basis' for the family shown here. Other combinations of these features might be relevant for specific applications (e.g., pulse width $T \delta $ instead of duty fraction $\delta$) but the principles behind the basis set of decoders should allow development of sensors for such features as well. We will assume that the amplitude of the input signal is fixed and will not consider amplitude fluctuations. A novel mechanism to convert oscillatory signals with such amplitude noise into a stereotyped signal was proposed recently\cite{Agrawal2015-bt}; such an amplitude buffering mechanism can be used upstream of the temporal decoders proposed here.

\begin{figure*}
    \centering
    \includegraphics{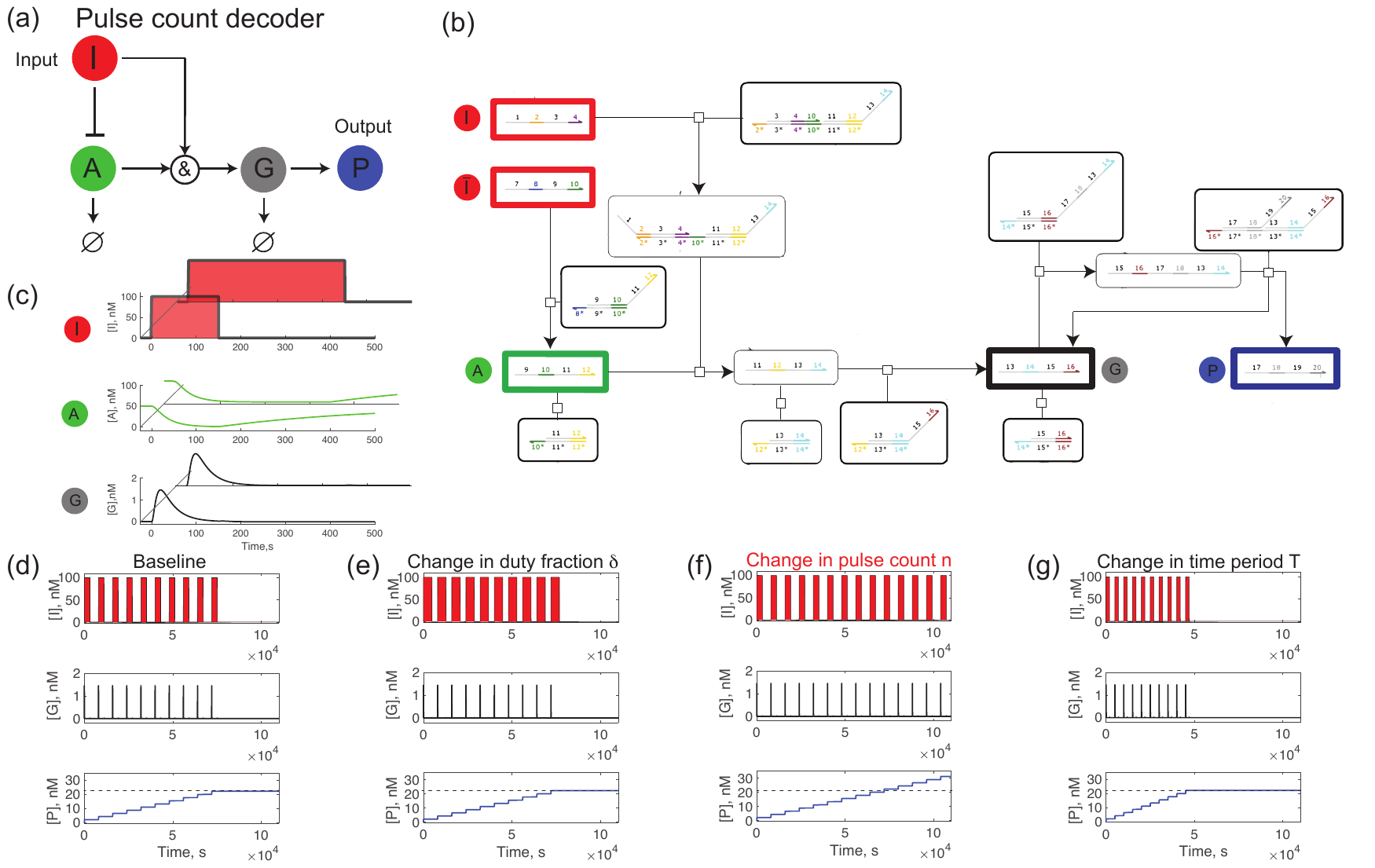}
    \caption{A pulse count decoder implemented by an Incoherent Feed-Forward Loop that responds only to step ups in input (but not step downs). (a) Reaction network to decode 
    pulse number. 
    Species $G$ is created only when $I$ and $A$ are present.  (b) DNA Strand Displacement (DSD) implementation of mechanism in (a) with waste products suppressed (see SI for full circuit). Species $\bar{I}$ is the negation of input $I$ as in dual-rail logic, i.e., $\bar{I}$ is high when ${I}$ is low and vice-versa.  
    (c) $G$ shows a stereotyped response to pulses of $I$ that is independent of the width and separation of such pulses in $I$. The stereotyped response is due to the `incoherent' regulation of $G$ by $I$; $I$ exerts a direct fast positive effect on $G$, causing a rapid rise, but also exerts an indirect delayed negative effect on $G$ by suppressing $A$ (green). (d) The output $P$ integrates and reports the number of stereotyped responses of $G$.  Consequently, changing the (e) duty fraction or (g) time period of pulses in $I$ has no impact on the output $P$ which does change with (f) pulse count.}
    \label{fig:pulsecount}
\end{figure*}

\subsection{Pulse number decoder}
We begin with a molecular circuit that can count the number of pulses $n$ seen, without regard to the width of each pulse or the separation between pulses. 

To count pulses $n$ in this manner, we first seek a circuit that produces a stereotyped response to each pulse that is independent of pulse width and separation. To do so, we take inspiration from `biochemical adaptation' mechanisms used e.g., in bacterial chemotaxis \cite{Tu2013-sk}. Such molecular circuits show a transient response to step \emph{changes} in an input signal but return to their prior state and are insensitive to the steady state value of the input. Incoherent Feed-Forward Loops (IFFL) and Negative Feedback loops are two common molecular circuits that carry out adaptation in biology\cite{FERRELL201662}. IFFLs have been recently implemented with DNA strand displacement reactions \cite{Chirieleison_2013}. Recent work on buffers with finite response time \cite{Scalise2018-jc} also resemble adaptation.

\begin{figure*}
    \centering
    \includegraphics{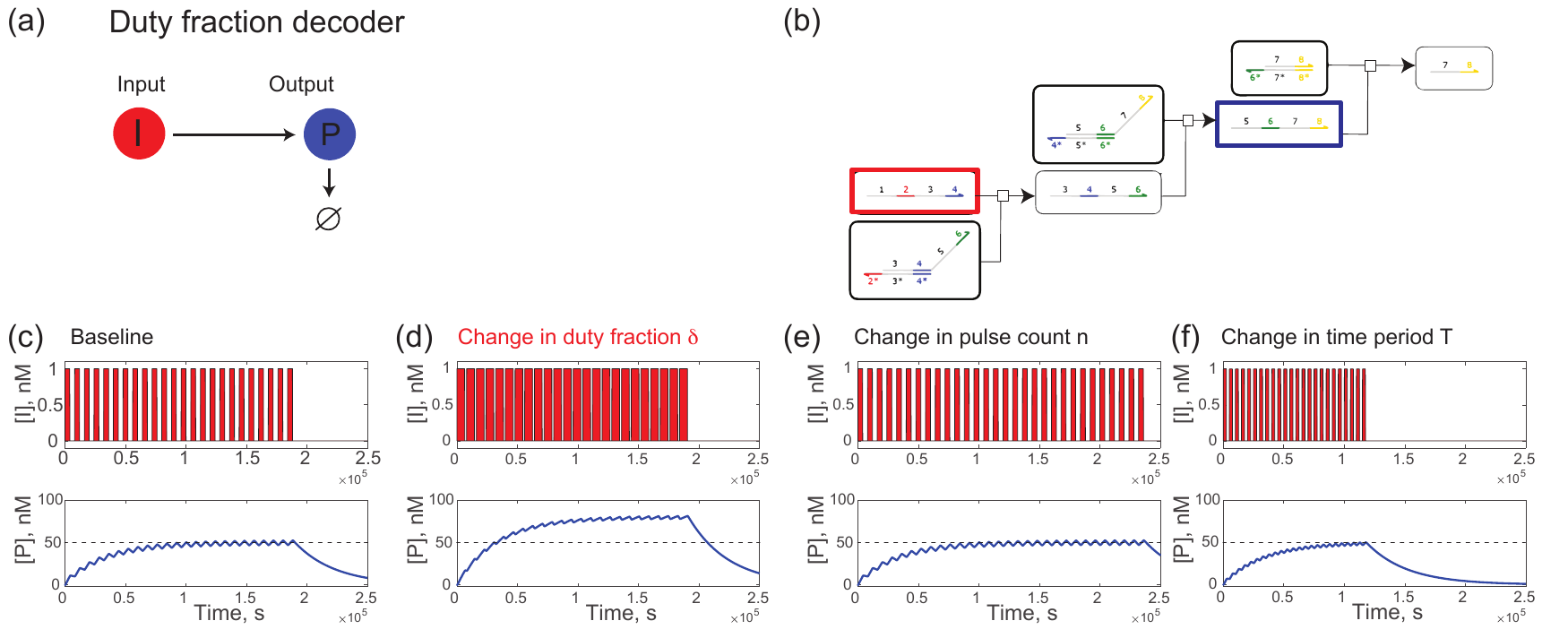}
    \caption{A duty fraction decoder implemented by a timed decay mechanism that computes a moving average. (a) Reaction network to decode duty fraction.   (b) DNA Strand Displacement (DSD) implementation of mechanism in (a). See Methods for kinetic parameters and SI for circuit that includes waste products. (c) Sample time traces for a typical input signal $[I]=c(t)$ from simulations of the DNA network. (d) Changing the duty fraction of input $[I]$ changes steady state value of output $[P]$. (e,f) Changing time period $T$ or pulse count $n$ do not affect the readout $[P]$.} 
    \label{fig:dutycycle}
\end{figure*}
\begin{figure*}
    \centering
    \includegraphics{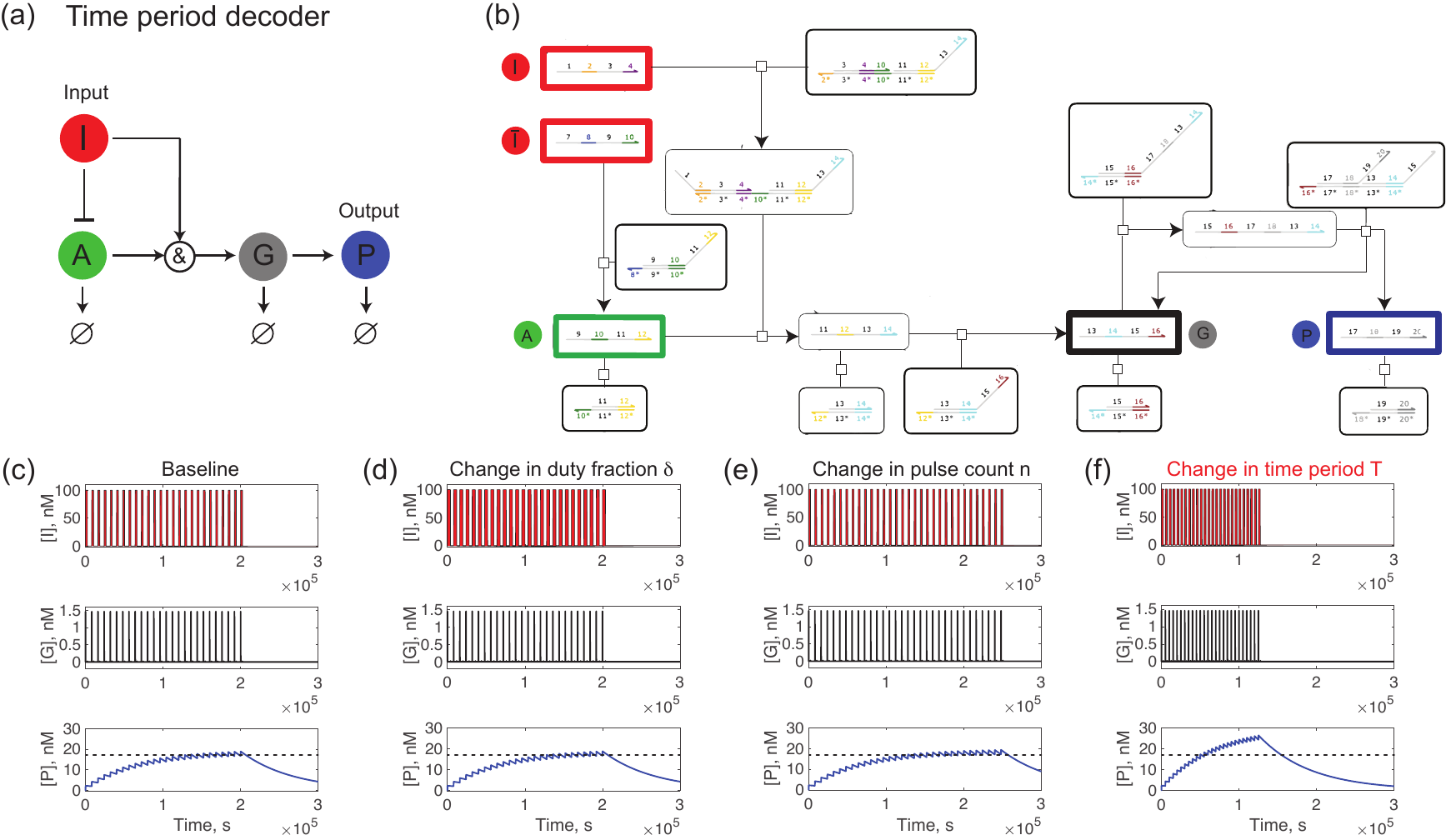}
    \caption{A time period decoder implemented by a combining an Incoherent Feed-Forward Loop with a timed decay. (a) Reaction network to decode 
    time period.  (b) DNA Strand Displacement (DSD) implementation of mechanism in (a) with waste products suppressed. Species $\bar{I}$ is the negation of input $I$ as in dual-rail logic.  (c) As with the pulse count decoder, $G$ shows a stereotyped response of fixed width $\tau_s$ to pulses of $I$.  However, the output $P$ now computes the moving average of the stereotyped responses of $G$ over a fixed timescale set by its decay constant. Hence the output $P$ reflects the duty fraction of the stereotyped pattern in $G$ which is $\tau_s/T$.  Consequently, changing the (d) duty fraction or (e) number of pulses in $I$ has no impact on the output $P$ which does change with (f) time period $T$.}
    \label{fig:timeperiod}
\end{figure*}

However, adaptive circuits in the biological literature often respond in an equal and opposite manner to both the rising and falling edges of each input pulse\cite{Tu2013-sk,Barkai1997-zq}. To count pulses, we desire an \emph{asymmetric} response to step ups and step downs in the input. For example, a circuit that responds only to step ups but ignores step downs and steady values of the input would naturally count pulses. Such asymmetric adaptation\cite{Mitchell2015-oa,Sorre2014-oc} can be naturally achieved in a CRN if the adaptive variable has a resting concentration at zero; then step downs can have little effect since concentrations cannot fall below zero.

Here, we present a simple CRN implementation of such an adaptive Incoherent Feed-Forward Loop circuit with an \emph{asymmetric} response to step ups and downs:
\begin{eqnarray}
\ce{I + A ->[k_1] I + G }\label{eq:IandA}\\
\ce{A  ->[\lambda_1] \phi }\\
\ce{\bar{I}   ->[k_2] A}\\
\ce{G ->[\lambda_2] \phi}\\
\ce{G ->[k_4] P + G}\label{eq:PfromG}
\end{eqnarray}
Here $\bar{I}$ refers to a species that is the negation of the input $I$ as in dual-rail logic\cite{Qian2011-Simple,Qian2011-yd}; i.e., we assume a second input species $\bar{I}$ that is high when the input $I$ is low and vice-versa. In principle, such dual rail $(\bar{I},I)$ input can be created from a single input $I$ by a fast reusable NOT-gate\cite{Qian2011-Simple}. One functional fast NOT-gate is provided in the Supplemental Information. 
\par
To understand the mechanism, consider the response to a single step up in $[I]$ shown in Fig.\ref{fig:pulsecount}c. Production of $G$ requires both $I$ and $A$ to be present; while $A$ is high in its resting state, $G$ starts being produced only when $I$ steps up. But meanwhile turning on $I$ also increases the degradation of $A$ on a timescale $1/(\lambda_1+k_1 [I])$, with $A$ eventually reaching a small steady state. As a result, $G$ stops being produced after a time $$\tau_{up}=\frac{\log{\frac{\lambda_2-\lambda_1}{k_1 [I]}}}{\lambda_2-k_1 [I]-\lambda_1}$$ 
and starts falling due to its own degradation timescale $1/\lambda_2$, thus producing a stereotyped response to the step up in $I$. This stereotyped response lasts a total time $\tau_a \sim \tau_{up}+1/\lambda_2$. A subsequent step down in $I$ that occurs more than a time $\tau_a$ after the step up has minimal impact on $G$ because $G$ is already near zero. After the step down of $I$, $A$ is restored back to its resting value on a timescale $1/\lambda_1$.  
If an output species $P$ is produced in response to $G$, the total amount of $P$ will report the number of pulses $n$ without regard to time period $T$ or duty fraction $\delta$.

\textit{Limits of operation:} There are two critical requirements for the mechanism above to work. First, the pulse width $T \delta $ must be larger than the length of the stereotyped adaptive response $\tau_a$, so that the stereotyped response is not interrupted by the step down in $I$. For our parameter choices,  $\lambda_2\approx k_1\gg \lambda_1$, we can approximate $\tau_{up}\approx{1/\lambda_2}$. Thus, $\tau_a \sim \tau_{up}+1/\lambda_2\sim 1/\lambda_2$ and we require that $ T \delta> 1/\lambda_2$. Second, the pulse off time $ T (1-\delta)$ needs to be long enough so that $A$ can be restored to its resting state before the next pulse in $I$ comes, hence requiring $T (1-\delta)  > 1/\lambda_1$. Taken together, we require $T > T_{min} = \sup(\{1/\lambda_1,1/\lambda_2\})$. In our CRN with $\lambda_1\ll \lambda_2$, $T_{min}=1/\lambda_1$. 

To verify that real chemical networks can operate in this kinetic regime, we designed a DNA strand displacement implementation of this CRN. As with all the DSD circuits we present here, our design process leaned on the reaction designs laid out in Soloveichik et al. \cite{DNAUniversal}.
A simplified representation of the reaction network with waste products suppressed is shown in Fig \ref{fig:pulsecount}b; see SI for the full network.  Simulating the DNA strand displacement network using Visual DSD software with realistic kinetic parameters, we find that the output is indeed sensitive to pulse number $n$ but insensitive to duty fraction $\delta$ and time period $T$ over a significant range.

\subsection{Duty fraction decoder}

We now develop a circuit needed to decode duty fraction, $\delta$. The Chemical Reaction Network capable of decoding $\delta$ is deceptively simple in topology,
\begin{eqnarray}
\ce{I ->[k] I + P}\\
\ce{P ->[\lambda] \phi}
\end{eqnarray}
where $I$ is the time-varying input species whose concentration changes over time as $[I] = c(t)$. Species $P$ is created by every pulse of $I$ but decays with a time constant $1/\lambda$. For any $c(t)$, this simple linear system has an output $P$ given by, 
\begin{eqnarray}
P(t) &=& k \int^{t}_{-\infty}  c(t') e^{-\lambda (t-t')}  dt'.
\end{eqnarray}

Intuitively, this operation effectively takes the moving average of input species $I$ over a time window $1/\lambda$. If $\lambda$ is in the right range, the output $P$ is insensitive to $T,n$. For a pulse train $c(t)$ that starts at $t=0$, we find that $A$ rises and oscillates about a steady state value; see Fig \ref{fig:dutycycle}. The mean level of $P$ goes as,
\begin{eqnarray}
    \bar{P} &= \left( \frac{e^{\lambda T \delta}-1}{e^{\lambda T}-1} \right) \; (1+e^{\lambda T (1-\delta)}-e^{-\lambda T n}-e^{-\lambda T(\delta+n)}).  
\end{eqnarray}
    
\textit{Limits of operation:} The above formula algebraically depends on all three features $n,T,$ and $\delta$. However, the $n$ dependence is exponentially suppressed if $\lambda n T \gg 1$. The $T$ dependence is also weak if $\lambda T \ll 1$. Under these two limiting operations, we arrive at an approximate expression for $\bar{P}$,
\begin{eqnarray}
    \bar{P} &\approx&  \left(\frac{k}{\lambda}\delta \right) ( 1 +  \frac{(1-3\delta+2\delta^2)}{12}(\lambda T)^2 + O((\lambda T)^3)) \nonumber \\ 
    &\approx& \frac{k}{\lambda}\delta 
\end{eqnarray}

Thus, the output is dependent only on the duty fraction $\delta$ and insensitive to $T,n$ provided $T < T_{max} = 1/\lambda$ and $n  > n_{min} = 1/(\lambda T)$.

Finally, note that the mean level of $P$ is a good readout only if the size of the oscillations about $\bar{P}$ seen in Fig.\ref{fig:dutycycle}c are small. The size of such oscillations relative to the mean is given by $\frac{\Delta \bar{P}}{\bar{P}} =  \lambda T (1-\delta) $ which is naturally small in the limits of operation defined above ($\lambda T \ll 1$ and $\delta \in [0,1]$). The operational regime can be expanded and oscillations further suppressed by appending an additional species $B$ with \ce{P -> P + B}, \ce{B -> \phi}. 

To check whether real chemical systems can operate in the kinetic regime defined above, we implemented this CRN using DNA strand displacement reactions.
We generated the network and equations using Visual DSD software \cite{Lakin2011-lb} with realistic kinetic parameters (see Methods/SI) and then simulated this system in MATLAB. The impact of varying $n,T,\delta$ on the output is shown in Fig. \ref{fig:dutycycle}c-f. As desired, the output $A(t)$ is only sensitive to changes in duty fraction $\delta$ and insensitive to changes in $n,T$.

\begin{figure*}
    \centering
    \includegraphics[width=\textwidth]{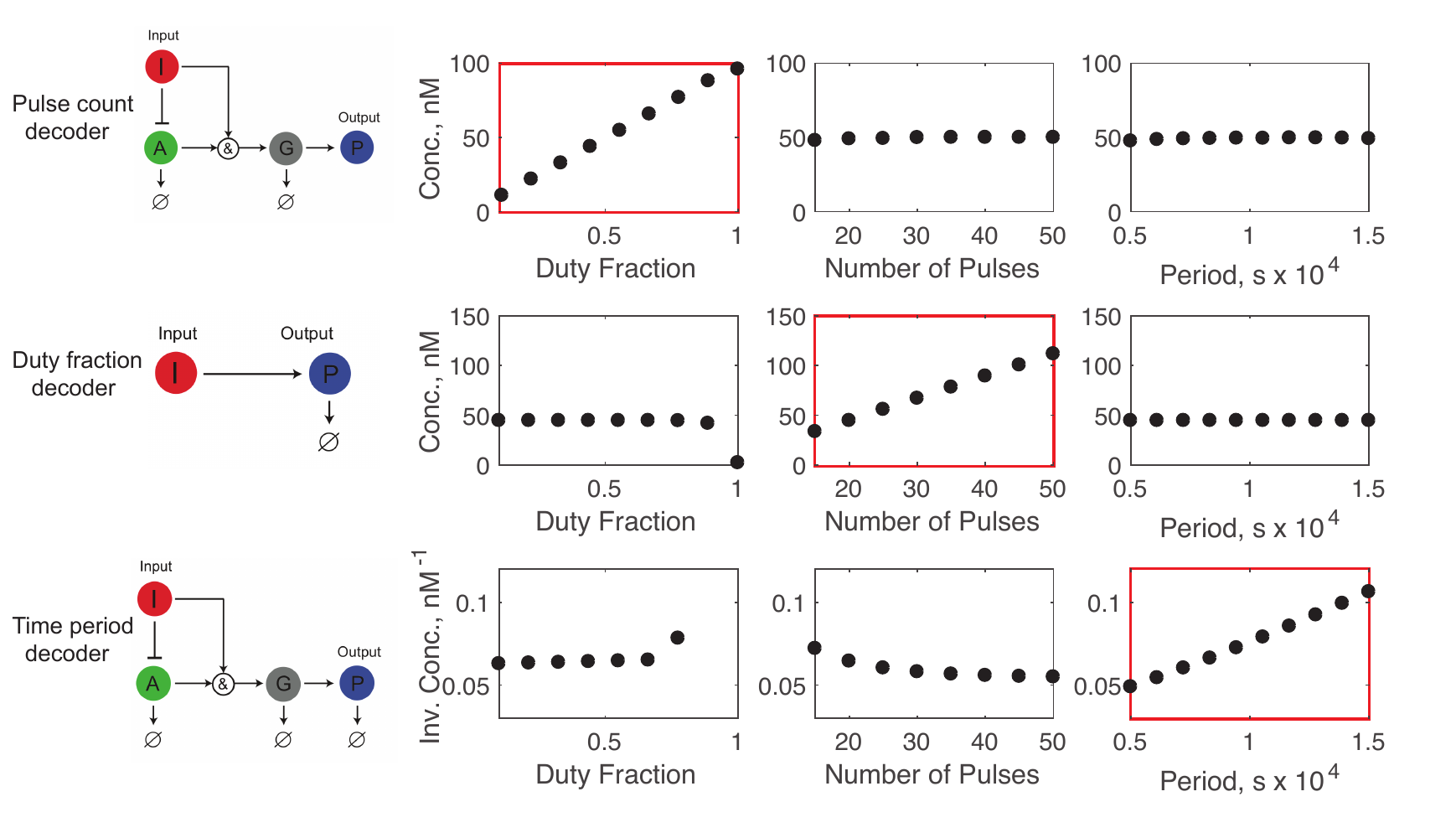}
    \caption{Output of DNA-based decoders with realistic kinetic parameters on a library of temporal patterns that systematically vary in pulse number, duty fraction and time period. We see that the output of each decoder circuit changes significantly in response to changes in one of the three features but is insensitive to the other two features. For example, the duty fraction decoder changes by a factor of $4 \times$ as duty fraction changes from $0.2$ to $0.8$ but only changes by a factor of $1.02$ as time period changes from $5000$ seconds to $15000$ seconds. The breakdown (e.g., high duty fraction for time period decoder) is described by the limits of operation derived in the text.  Thus, taken together, the decoders demonstrated here, with realistic DNA hybridization-based reaction rates, can distinguish temporal patterns accurately over a range of timescales.}
    \label{fig:fulldecoder}
\end{figure*}
\subsection{Time period decoder}

We can build a time period $T$ decoder by modifying the adaptive circuit motif introduced above; we simply add a decay process for species $P$ so that $[P]$ reflects the moving average of $G$ over a fixed timescale $1/\lambda_3$,
\begin{eqnarray}
\ce{P ->[\lambda_3] \phi}.
\end{eqnarray}

All other reactions for $I,A,G,P$ are as shown in Eqn.\ref{eq:IandA} - \ref{eq:PfromG}. 
The analysis for this CRN looks identical to that done for our duty fraction sensor. However, now we are taking the moving average of $[G](t)$ which possesses the same periodicity $T$ as the initial pulse train but has pulses with a duration set only by the kinetic parameters of the network $\tau$.  Thus, $[G](t)$ has an effective duty fraction $\delta_{eff}=\tau/T$. If we tune kinetic parameters to the same regime from our duty fraction sensor, the output species $P$ will then report this $\delta_{eff}$ (up to geometric factor for the shape of the pulse) and is thus proportional to $1/T$ but independent of the $\delta$ and $n$ from the initial species $I$.

\textit{Limits of operation:} This CRN directly inherits the kinetic constraints from both the pulse number decoder applied to the $I,A,G$ part of the network and the duty fraction decoder applied to $G,P$ part of the network. The former restricts $T (1-\delta) > 1/\lambda_1$ as discussed for the pulse number decoder. The latter constraint requires $T < 1/\lambda_3$ and $ n > 1/(\lambda_3 T)$ as discussed for the duty fraction decoder. In this kinetic regime, our output depends strongly on $T$ and only weakly on $n,\delta$.

The DNA implementation is shown in Fig \ref{fig:timeperiod}b.
By choosing kinetic constants listed in Methods, we simulated this network and see that the output is insensitive to changes in $\delta,n$ but sensitive to changes in $T$.

Finally, we systematically tested all three of the mechanisms proposed here against a library of temporal patterns that vary in all three features ($n,\delta,T$). Each decoder was implemented with DNA strand displacement with fixed kinetic rates.  See Fig.\ref{fig:fulldecoder}. We see that each decoder shows a much larger response to changes in its relevant feature than to the other features over a substantial dynamic range. Thus, collectively, the three circuits can discriminate each member of the temporal pattern library.

\section{Discussion}
The circuits introduced here exploit a weakness in the context of digital computation - internal transients - to naturally perform analog computation on temporally coded information. The performance of such analog `computation using transients' is limited by the dynamic range over which timescales of transients can be tuned. Hence DNA strand displacement reactions is particularly suitable for such computation since their kinetics can be tuned over a large dynamic range through toehold sequence design \cite{Soloveichik2010-kz,Zhang2011-by,Machinek2014-di,Ouldridge2015}.

The methods developed here can be combined with other developments in the molecular technology field\cite{Benzinger2018-pf}. For example, a drug payload carried by a DNA origami pill\cite{LINKO2015586} can be released only in those cells with a pulsatile pattern of the transcription factor NFkB that precedes a inflammatory response but not in cells with NFkB patterns that precede an adaptive immune response\cite{Purvis2013-io}. Similar \emph{in situ} temporal computation using molecules can also help surveil complex ecosystems, such as the gut, where a future ecological collapse is often indicated by temporal precursors \cite{Dai1175}. 

We have produced circuits that decode broadly relevant but \emph{predetermined} temporal features. Going forward, it would be interesting to develop molecular circuits that can \emph{learn} relevant temporal features dynamically \cite{Lakin2016-jv, Lakin2014-qp,Poole2017-cz} as in machine learning approaches.
In the learning paradigm, for example, a molecular circuit could be exposed to two classes of time-varying patterns during a `training phase' (the temporal equivalent of cat and dog images in static pattern recognition). The circuit would determine which temporal features can best distinguish those two classes. 

%

\begin{acknowledgments}

The authors thank Aaron Dinner, Anders Hansen, Kabir Husain, John Hopfield, Sidney Nagel and Michael Rust for useful discussions. We acknowledge NSF-MRSEC 1420709 for funding and the University of Chicago Research Computing Center for computing resources. AM is grateful to the Simons Foundation's Mathematical Modeling of Living Systems investigator program for support.

\end{acknowledgments}

\appendix

\section{Methods}

We first formulated abstract chemical networks with the desired feature detecting properties. Then, largely following the design principles laid out in \cite{DNAUniversal}, we implemented these CRNs as DNA strand displacement reactions.
All strand displacement circuits are designed within the Visual DSD software described in \cite{Lakin2011-lb} using default kinetic parameters (unless explicitly noted in the Supplemental Information) and concentrations $\in [.05 nM, 1 mM]$. Then, by adapting the MATLAB code generated within this program, we exposed these circuits to the pulsatile inputs defined in the main text, defined by their duty fraction $\delta$, number of pulses $n$, and period $T$. All results shown are from deterministic simulations. 
\par 
\section{Detailed Chemical Networks}
All reactions shown utilize the default kinetic parameters within Visual DSD ($3\times 10^{-4} (nM s)^{-1}$ bind, $.1226 s^{-1}$ unbind corresponding to toe-holds with 4-6 nucleotides \cite{ControlofDNAWinfree}). Different reaction rates were largely achieved by selecting appropriate initial concentrations. Species with specified initial concentrations are outlined in bold and their values are given in accompanying tables. All species whose initial concentrations are specified and do not have explicit time dependence displayed in the main text are held at their initial concentrations by hand.

\begin{figure*}[h]
    \centering
    \includegraphics[scale=1]{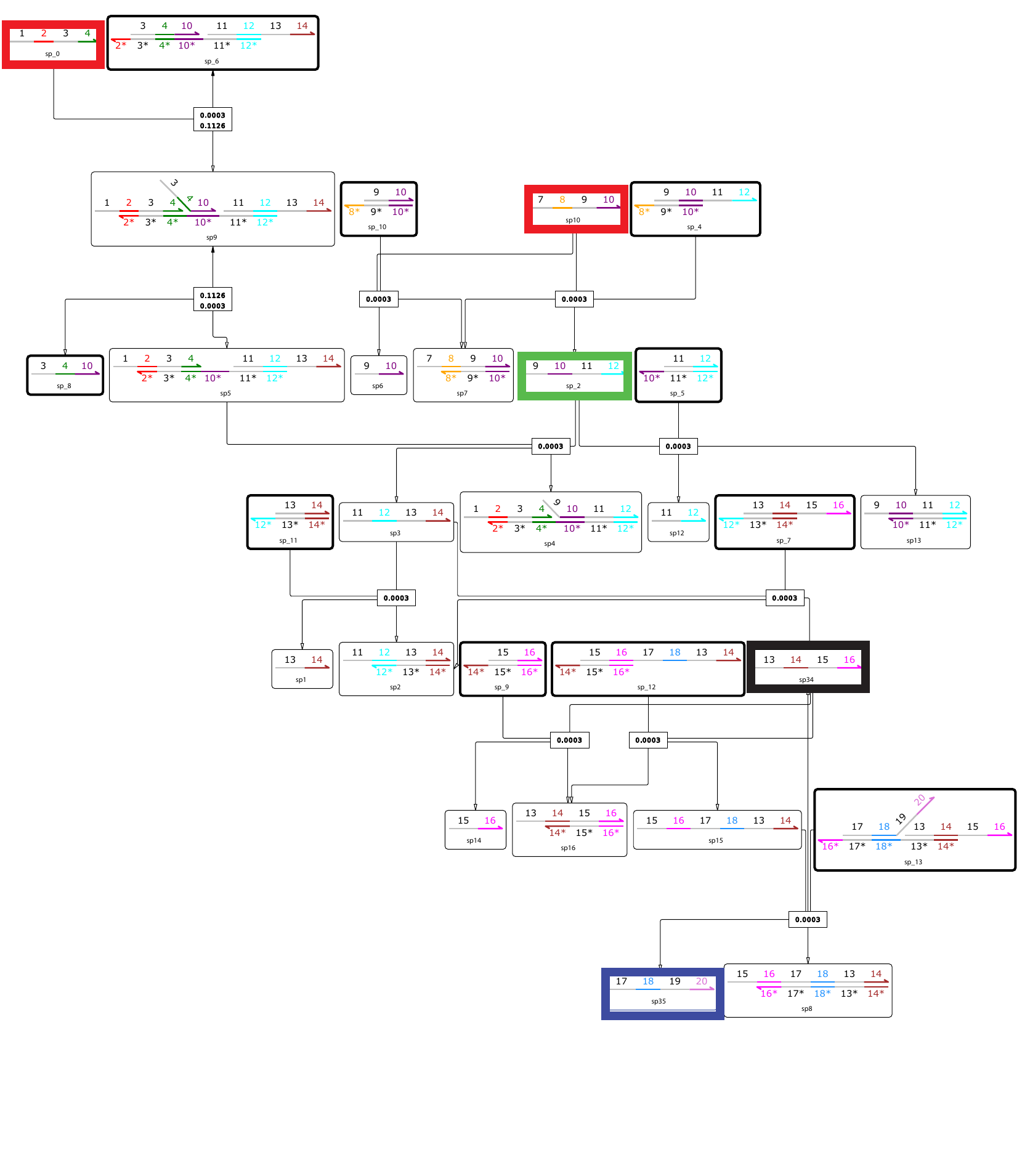}
    \caption{Pulse count decoder. Here we present the full DSD reaction network for our pulse counting circuit. See the main text for an analysis of its dynamics (Fig. 2) and performance (Fig. 5). The two species in red had their dynamics directly modulated to the parameters of the input series, with sp10 pulsing exactly out of phase with sp\_6. Graphs and labels are generated automatically withing the Visual DSD software \cite{Lakin2011-lb}. See Table \ref{tab:counttable} for a list of initial concentrations.}
    \label{fig:SItimerperiod}
\end{figure*}

\begin{figure*}[h]
    \centering
    \includegraphics{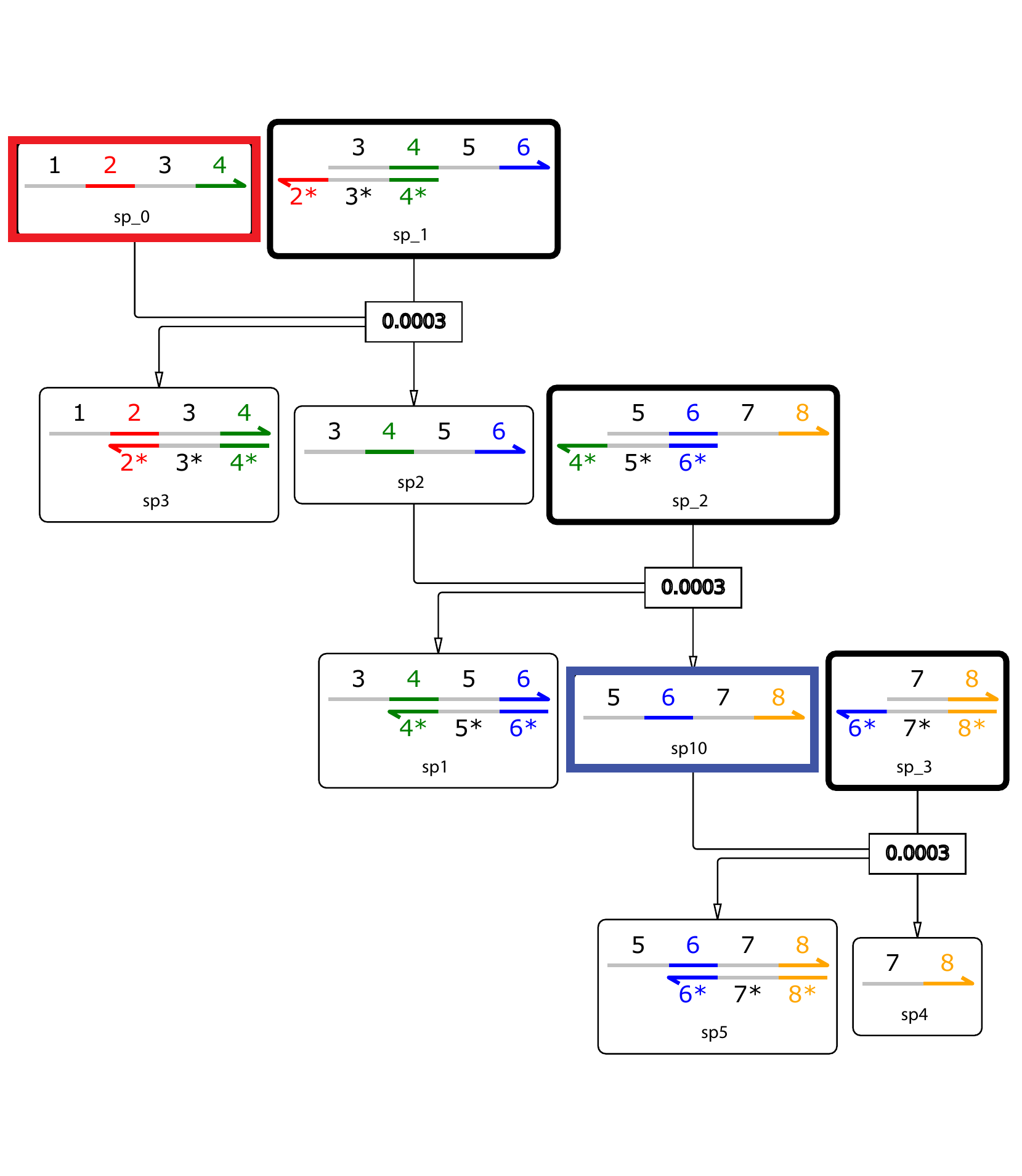}
    \caption{Duty fraction decoder. This circuit effectively take the moving average of the dynamics of sp\_0 and reports it in sp10. See main text Figure 3 for analysis and Figure 5 for performance. Initial concentrations of bolded species are listed in Table \ref{tab:dutytable}}
    \label{fig:SIdutyfraction}
\end{figure*}

\begin{figure*}[h]
    \centering
    \includegraphics{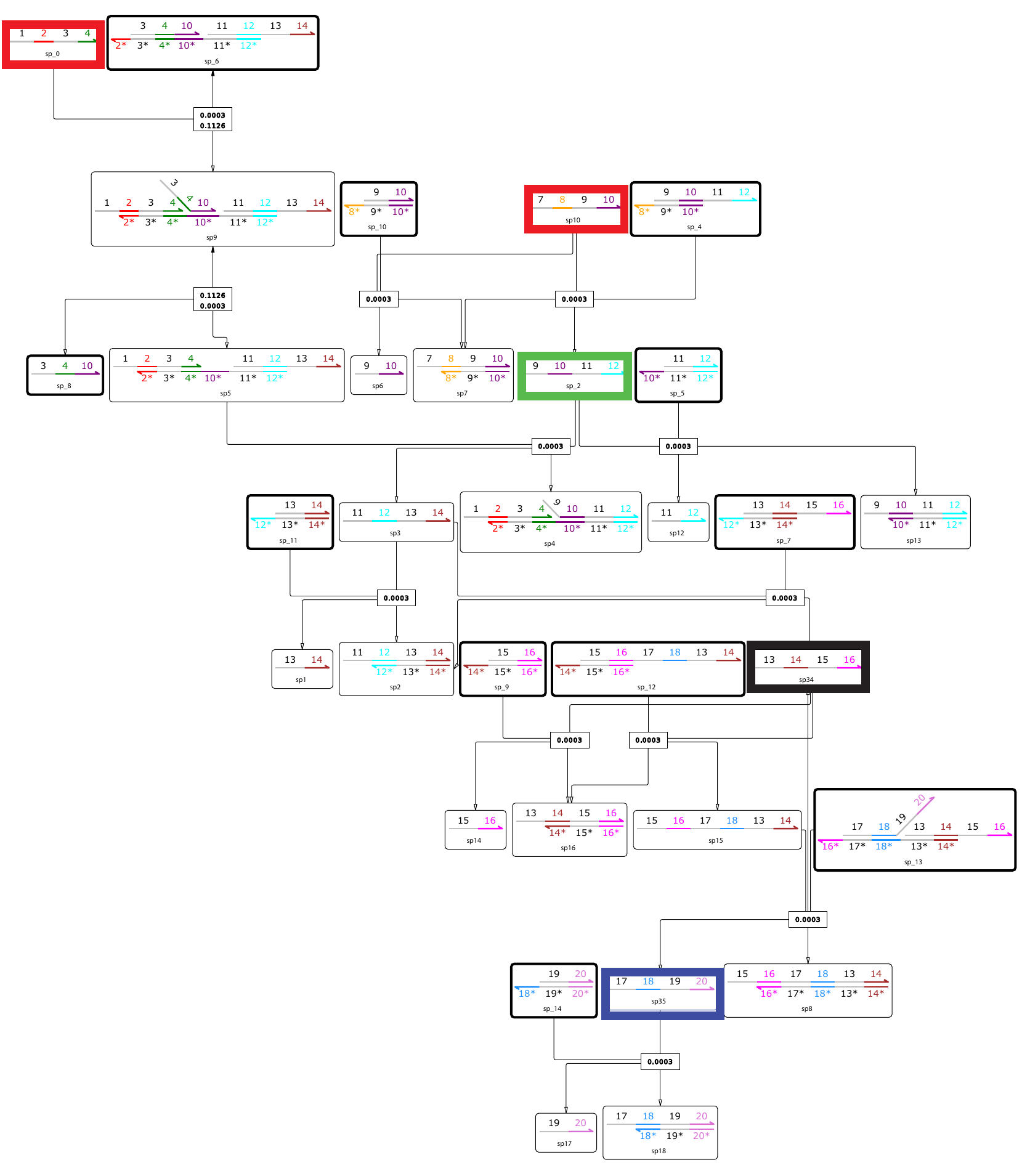}
    \caption{Time period decode. By taking the moving average of sp34, this circuit decodes the period of sp\_0. See main text Figure 4 for analysis of dynamics and Figure 5 for analysis of performance. Initial concentrations of bolded species are reported in Table \ref{tab:periodtable}.}
    \label{fig:SItimeperiod}
\end{figure*}

\begin{table}
\caption{Pulse Counter Initial Conditions} 
\centering 
\begin{tabular}{c c c c} 
\hline\hline 
Species & Initial Conc. (nM) \\ [0.5ex] 
\hline 
sp\_0 & 100 (0)\\ 
sp\_6 & 10000 \\
sp\_2 & 10  \\
sp\_10 & 10000  \\
sp10 & 100 (0)  \\
sp\_5 & 100 \\
sp\_7 & 10000 \\
sp\_8 & 10000 \\
sp\_11 & 10000 \\
sp\_12 & 100 \\
sp\_13 & 1000 \\
\hline 
\end{tabular}
\label{tab:counttable} 
\end{table}

\begin{table}
\caption{Duty Fraction Sensor Initial Conditions} 
\centering 
\begin{tabular}{c c c c} 
\hline\hline 
Species & Initial Conc. (nM) \\ [0.5ex] 
sp\_0 & 1000(0) \\
sp\_1 & 100 \\
sp\_2 & 100 \\
sp\_3 & 10 \\
\hline 
\end{tabular}
\label{tab:dutytable} 
\end{table}

\begin{table}
\caption{Period Sensor Initial Conditions} 
\centering 
\begin{tabular}{c c c c} 
\hline\hline 
Species & Initial Conc. (nM) \\ [0.5ex] 
\hline 
sp\_0 & 100 (0)\\ 
sp\_6 & 10000 \\
sp\_2 & 10  \\
sp\_10 & 10000  \\
sp10 & 100 (0)  \\
sp\_5 & 100 \\
sp\_7 & 10000 \\
sp\_8 & 10000 \\
sp\_11 & 10000 \\
sp\_12 & 100 \\
sp\_13 & 1000 \\
sp\_14 & .05 \\
\hline 
\end{tabular}
\label{tab:periodtable} 
\end{table}


\bibliography{Paperpile,additionalCitations}

\begin{thebibliography}{53}%
\makeatletter
\providecommand \@ifxundefined [1]{%
 \@ifx{#1\undefined}
}%
\providecommand \@ifnum [1]{%
 \ifnum #1\expandafter \@firstoftwo
 \else \expandafter \@secondoftwo
 \fi
}%
\providecommand \@ifx [1]{%
 \ifx #1\expandafter \@firstoftwo
 \else \expandafter \@secondoftwo
 \fi
}%
\providecommand \natexlab [1]{#1}%
\providecommand \enquote  [1]{``#1''}%
\providecommand \bibnamefont  [1]{#1}%
\providecommand \bibfnamefont [1]{#1}%
\providecommand \citenamefont [1]{#1}%
\providecommand \href@noop [0]{\@secondoftwo}%
\providecommand \href [0]{\begingroup \@sanitize@url \@href}%
\providecommand \@href[1]{\@@startlink{#1}\@@href}%
\providecommand \@@href[1]{\endgroup#1\@@endlink}%
\providecommand \@sanitize@url [0]{\catcode `\\12\catcode `\$12\catcode
  `\&12\catcode `\#12\catcode `\^12\catcode `\_12\catcode `\%12\relax}%
\providecommand \@@startlink[1]{}%
\providecommand \@@endlink[0]{}%
\providecommand \url  [0]{\begingroup\@sanitize@url \@url }%
\providecommand \@url [1]{\endgroup\@href {#1}{\urlprefix }}%
\providecommand \urlprefix  [0]{URL }%
\providecommand \Eprint [0]{\href }%
\providecommand \doibase [0]{http://dx.doi.org/}%
\providecommand \selectlanguage [0]{\@gobble}%
\providecommand \bibinfo  [0]{\@secondoftwo}%
\providecommand \bibfield  [0]{\@secondoftwo}%
\providecommand \translation [1]{[#1]}%
\providecommand \BibitemOpen [0]{}%
\providecommand \bibitemStop [0]{}%
\providecommand \bibitemNoStop [0]{.\EOS\space}%
\providecommand \EOS [0]{\spacefactor3000\relax}%
\providecommand \BibitemShut  [1]{\csname bibitem#1\endcsname}%
\let\auto@bib@innerbib\@empty
\bibitem [{\citenamefont {Song}\ \emph {et~al.}(2008)\citenamefont {Song},
  \citenamefont {Wang}, \citenamefont {Li}, \citenamefont {Fan},\ and\
  \citenamefont {Zhao}}]{Song2008-tv}%
  \BibitemOpen
  \bibfield  {author} {\bibinfo {author} {\bibfnamefont {S.}~\bibnamefont
  {Song}}, \bibinfo {author} {\bibfnamefont {L.}~\bibnamefont {Wang}}, \bibinfo
  {author} {\bibfnamefont {J.}~\bibnamefont {Li}}, \bibinfo {author}
  {\bibfnamefont {C.}~\bibnamefont {Fan}}, \ and\ \bibinfo {author}
  {\bibfnamefont {J.}~\bibnamefont {Zhao}},\ }\href {\doibase
  10.1016/j.trac.2007.12.004} {\bibfield  {journal} {\bibinfo  {journal}
  {Trends Analyt. Chem.}\ }\textbf {\bibinfo {volume} {27}},\ \bibinfo {pages}
  {108} (\bibinfo {year} {2008})}\BibitemShut {NoStop}%
\bibitem [{\citenamefont {Modi}\ \emph {et~al.}(2009)\citenamefont {Modi},
  \citenamefont {M~G}, \citenamefont {Goswami}, \citenamefont {Gupta},
  \citenamefont {Mayor},\ and\ \citenamefont {Krishnan}}]{Modi2009-li}%
  \BibitemOpen
  \bibfield  {author} {\bibinfo {author} {\bibfnamefont {S.}~\bibnamefont
  {Modi}}, \bibinfo {author} {\bibfnamefont {S.}~\bibnamefont {M~G}}, \bibinfo
  {author} {\bibfnamefont {D.}~\bibnamefont {Goswami}}, \bibinfo {author}
  {\bibfnamefont {G.~D.}\ \bibnamefont {Gupta}}, \bibinfo {author}
  {\bibfnamefont {S.}~\bibnamefont {Mayor}}, \ and\ \bibinfo {author}
  {\bibfnamefont {Y.}~\bibnamefont {Krishnan}},\ }\href {\doibase
  10.1038/nnano.2009.83} {\bibfield  {journal} {\bibinfo  {journal} {Nat.
  Nanotechnol.}\ }\textbf {\bibinfo {volume} {4}},\ \bibinfo {pages} {325}
  (\bibinfo {year} {2009})}\BibitemShut {NoStop}%
\bibitem [{\citenamefont {Spichiger-Keller}(2008)}]{Spichiger-Keller2008-qr}%
  \BibitemOpen
  \bibfield  {author} {\bibinfo {author} {\bibfnamefont {U.~E.}\ \bibnamefont
  {Spichiger-Keller}},\ }\href@noop {} {{\emph {\bibinfo
  {title} {Chemical Sensors and Biosensors for Medical and Biological
  Applications}}}}\ (\bibinfo  {publisher} {John Wiley and Sons},\ \bibinfo
  {year} {2008})\BibitemShut {NoStop}%
\bibitem [{\citenamefont {Chen}\ \emph {et~al.}(2013)\citenamefont {Chen},
  \citenamefont {Briggs}, \citenamefont {McLain},\ and\ \citenamefont
  {Ellington}}]{Chen2013-ry}%
  \BibitemOpen
  \bibfield  {author} {\bibinfo {author} {\bibfnamefont {X.}~\bibnamefont
  {Chen}}, \bibinfo {author} {\bibfnamefont {N.}~\bibnamefont {Briggs}},
  \bibinfo {author} {\bibfnamefont {J.~R.}\ \bibnamefont {McLain}}, \ and\
  \bibinfo {author} {\bibfnamefont {A.~D.}\ \bibnamefont {Ellington}},\ }\href
  {\doibase 10.1073/pnas.1222807110} {\bibfield  {journal} {\bibinfo  {journal}
  {Proc. Natl. Acad. Sci. U. S. A.}\ }\textbf {\bibinfo {volume} {110}},\
  \bibinfo {pages} {5386} (\bibinfo {year} {2013})}\BibitemShut {NoStop}%
\bibitem [{\citenamefont {Purvis}\ and\ \citenamefont
  {Lahav}(2013)}]{Purvis2013-io}%
  \BibitemOpen
  \bibfield  {author} {\bibinfo {author} {\bibfnamefont {J.~E.}\ \bibnamefont
  {Purvis}}\ and\ \bibinfo {author} {\bibfnamefont {G.}~\bibnamefont {Lahav}},\
  }\href@noop {} {\bibfield  {journal} {\bibinfo  {journal} {Cell}\ }\textbf
  {\bibinfo {volume} {152}},\ \bibinfo {pages} {945} (\bibinfo {year}
  {2013})}\BibitemShut {NoStop}%
\bibitem [{\citenamefont {Hao}\ and\ \citenamefont
  {O'Shea}(2011)}]{Hao2011-fr}%
  \BibitemOpen
  \bibfield  {author} {\bibinfo {author} {\bibfnamefont {N.}~\bibnamefont
  {Hao}}\ and\ \bibinfo {author} {\bibfnamefont {E.~K.}\ \bibnamefont
  {O'Shea}},\ }\href {\doibase 10.1038/nsmb.2192} {\bibfield  {journal}
  {\bibinfo  {journal} {Nat. Struct. Mol. Biol.}\ }\textbf {\bibinfo {volume}
  {19}},\ \bibinfo {pages} {31} (\bibinfo {year} {2011})}\BibitemShut {NoStop}%
\bibitem [{\citenamefont {Hansen}\ and\ \citenamefont
  {O'Shea}(2013)}]{Hansen2013-cs}%
  \BibitemOpen
  \bibfield  {author} {\bibinfo {author} {\bibfnamefont {A.~S.}\ \bibnamefont
  {Hansen}}\ and\ \bibinfo {author} {\bibfnamefont {E.~K.}\ \bibnamefont
  {O'Shea}},\ }\href {\doibase 10.1038/msb.2013.56} {\bibfield  {journal}
  {\bibinfo  {journal} {Mol. Syst. Biol.}\ }\textbf {\bibinfo {volume} {9}},\
  \bibinfo {pages} {704} (\bibinfo {year} {2013})}\BibitemShut {NoStop}%
\bibitem [{\citenamefont {Hansen}\ and\ \citenamefont
  {O'Shea}(2016)}]{Hansen2016-fo}%
  \BibitemOpen
  \bibfield  {author} {\bibinfo {author} {\bibfnamefont {A.~S.}\ \bibnamefont
  {Hansen}}\ and\ \bibinfo {author} {\bibfnamefont {E.~K.}\ \bibnamefont
  {O'Shea}},\ }\href {\doibase 10.1016/j.cub.2016.02.058} {\bibfield  {journal}
  {\bibinfo  {journal} {Curr. Biol.}\ }\textbf {\bibinfo {volume} {26}},\
  \bibinfo {pages} {R269} (\bibinfo {year} {2016})}\BibitemShut {NoStop}%
\bibitem [{\citenamefont {Lakin}\ \emph {et~al.}(2011)\citenamefont {Lakin},
  \citenamefont {Youssef}, \citenamefont {Polo}, \citenamefont {Emmott},\ and\
  \citenamefont {Phillips}}]{Lakin2011-lb}%
  \BibitemOpen
  \bibfield  {author} {\bibinfo {author} {\bibfnamefont {M.~R.}\ \bibnamefont
  {Lakin}}, \bibinfo {author} {\bibfnamefont {S.}~\bibnamefont {Youssef}},
  \bibinfo {author} {\bibfnamefont {F.}~\bibnamefont {Polo}}, \bibinfo {author}
  {\bibfnamefont {S.}~\bibnamefont {Emmott}}, \ and\ \bibinfo {author}
  {\bibfnamefont {A.}~\bibnamefont {Phillips}},\ }\href {\doibase
  10.1093/bioinformatics/btr543} {\bibfield  {journal} {\bibinfo  {journal}
  {Bioinformatics}\ }\textbf {\bibinfo {volume} {27}},\ \bibinfo {pages} {3211}
  (\bibinfo {year} {2011})}\BibitemShut {NoStop}%
\bibitem [{\citenamefont {Del~Grosso}\ \emph {et~al.}(2015)\citenamefont
  {Del~Grosso}, \citenamefont {Dallaire}, \citenamefont
  {Vall{\'e}e-B{\'e}lisle},\ and\ \citenamefont {Ricci}}]{Del_Grosso2015-ce}%
  \BibitemOpen
  \bibfield  {author} {\bibinfo {author} {\bibfnamefont {E.}~\bibnamefont
  {Del~Grosso}}, \bibinfo {author} {\bibfnamefont {A.-M.}\ \bibnamefont
  {Dallaire}}, \bibinfo {author} {\bibfnamefont {A.}~\bibnamefont
  {Vall{\'e}e-B{\'e}lisle}}, \ and\ \bibinfo {author} {\bibfnamefont
  {F.}~\bibnamefont {Ricci}},\ }\href {\doibase 10.1021/acs.nanolett.5b04566}
  {\bibfield  {journal} {\bibinfo  {journal} {Nano Lett.}\ }\textbf {\bibinfo
  {volume} {15}},\ \bibinfo {pages} {8407} (\bibinfo {year}
  {2015})}\BibitemShut {NoStop}%
\bibitem [{\citenamefont {Kim}\ and\ \citenamefont
  {Winfree}(2011)}]{Kim2011-bz}%
  \BibitemOpen
  \bibfield  {author} {\bibinfo {author} {\bibfnamefont {J.}~\bibnamefont
  {Kim}}\ and\ \bibinfo {author} {\bibfnamefont {E.}~\bibnamefont {Winfree}},\
  }\href {\doibase 10.1038/msb.2010.119} {\bibfield  {journal} {\bibinfo
  {journal} {Mol. Syst. Biol.}\ }\textbf {\bibinfo {volume} {7}},\ \bibinfo
  {pages} {465} (\bibinfo {year} {2011})}\BibitemShut {NoStop}%
\bibitem [{\citenamefont {Qian}\ and\ \citenamefont
  {Winfree}(2009)}]{Qian2009-lq}%
  \BibitemOpen
  \bibfield  {author} {\bibinfo {author} {\bibfnamefont {L.}~\bibnamefont
  {Qian}}\ and\ \bibinfo {author} {\bibfnamefont {E.}~\bibnamefont {Winfree}},\
  }\href@noop {} {\bibfield  {journal} {\bibinfo  {journal} {DNA Computing}\ ,\
  \bibinfo {pages} {70}} (\bibinfo {year} {2009})}\BibitemShut {NoStop}%
\bibitem [{\citenamefont {Qian}\ and\ \citenamefont
  {Winfree}(2011)}]{Qian2011-yd}%
  \BibitemOpen
  \bibfield  {author} {\bibinfo {author} {\bibfnamefont {L.}~\bibnamefont
  {Qian}}\ and\ \bibinfo {author} {\bibfnamefont {E.}~\bibnamefont {Winfree}},\
  }\href@noop {} {\enquote {\bibinfo {title} {Scaling up digital circuit
  computation with {DNA} strand displacement cascades},}\ } (\bibinfo {year}
  {2011})\BibitemShut {NoStop}%
\bibitem [{\citenamefont {Qian}\ \emph {et~al.}(2011)\citenamefont {Qian},
  \citenamefont {Winfree},\ and\ \citenamefont {Bruck}}]{Qian2011-jh}%
  \BibitemOpen
  \bibfield  {author} {\bibinfo {author} {\bibfnamefont {L.}~\bibnamefont
  {Qian}}, \bibinfo {author} {\bibfnamefont {E.}~\bibnamefont {Winfree}}, \
  and\ \bibinfo {author} {\bibfnamefont {J.}~\bibnamefont {Bruck}},\ }\href
  {\doibase 10.1038/nature10262} {\bibfield  {journal} {\bibinfo  {journal}
  {Nature}\ }\textbf {\bibinfo {volume} {475}},\ \bibinfo {pages} {368}
  (\bibinfo {year} {2011})}\BibitemShut {NoStop}%
\bibitem [{\citenamefont {Cherry}\ and\ \citenamefont
  {Qian}(2018)}]{Cherry2018-qf}%
  \BibitemOpen
  \bibfield  {author} {\bibinfo {author} {\bibfnamefont {K.~M.}\ \bibnamefont
  {Cherry}}\ and\ \bibinfo {author} {\bibfnamefont {L.}~\bibnamefont {Qian}},\
  }\href {\doibase 10.1038/s41586-018-0289-6} {\bibfield  {journal} {\bibinfo
  {journal} {Nature}\ }\textbf {\bibinfo {volume} {559}},\ \bibinfo {pages}
  {370} (\bibinfo {year} {2018})}\BibitemShut {NoStop}%
\bibitem [{\citenamefont {Daniel}\ \emph {et~al.}(2013)\citenamefont {Daniel},
  \citenamefont {Rubens}, \citenamefont {Sarpeshkar},\ and\ \citenamefont
  {Lu}}]{Daniel2013-sn}%
  \BibitemOpen
  \bibfield  {author} {\bibinfo {author} {\bibfnamefont {R.}~\bibnamefont
  {Daniel}}, \bibinfo {author} {\bibfnamefont {J.~R.}\ \bibnamefont {Rubens}},
  \bibinfo {author} {\bibfnamefont {R.}~\bibnamefont {Sarpeshkar}}, \ and\
  \bibinfo {author} {\bibfnamefont {T.~K.}\ \bibnamefont {Lu}},\ }\href
  {\doibase 10.1038/nature12148} {\bibfield  {journal} {\bibinfo  {journal}
  {Nature}\ }\textbf {\bibinfo {volume} {497}},\ \bibinfo {pages} {619}
  (\bibinfo {year} {2013})}\BibitemShut {NoStop}%
\bibitem [{\citenamefont {Song}\ \emph {et~al.}(2018)\citenamefont {Song},
  \citenamefont {Garg}, \citenamefont {Mokhtar}, \citenamefont {Bui},\ and\
  \citenamefont {Reif}}]{Song2018-ba}%
  \BibitemOpen
  \bibfield  {author} {\bibinfo {author} {\bibfnamefont {T.}~\bibnamefont
  {Song}}, \bibinfo {author} {\bibfnamefont {S.}~\bibnamefont {Garg}}, \bibinfo
  {author} {\bibfnamefont {R.}~\bibnamefont {Mokhtar}}, \bibinfo {author}
  {\bibfnamefont {H.}~\bibnamefont {Bui}}, \ and\ \bibinfo {author}
  {\bibfnamefont {J.}~\bibnamefont {Reif}},\ }\href {\doibase
  10.1021/acssynbio.6b00390} {\bibfield  {journal} {\bibinfo  {journal} {ACS
  Synth. Biol.}\ }\textbf {\bibinfo {volume} {7}},\ \bibinfo {pages} {46}
  (\bibinfo {year} {2018})}\BibitemShut {NoStop}%
\bibitem [{\citenamefont {Song}\ \emph {et~al.}(2016)\citenamefont {Song},
  \citenamefont {Garg}, \citenamefont {Mokhtar}, \citenamefont {Bui},\ and\
  \citenamefont {Reif}}]{Song2016-ex}%
  \BibitemOpen
  \bibfield  {author} {\bibinfo {author} {\bibfnamefont {T.}~\bibnamefont
  {Song}}, \bibinfo {author} {\bibfnamefont {S.}~\bibnamefont {Garg}}, \bibinfo
  {author} {\bibfnamefont {R.}~\bibnamefont {Mokhtar}}, \bibinfo {author}
  {\bibfnamefont {H.}~\bibnamefont {Bui}}, \ and\ \bibinfo {author}
  {\bibfnamefont {J.}~\bibnamefont {Reif}},\ }\href {\doibase
  10.1021/acssynbio.6b00144} {\bibfield  {journal} {\bibinfo  {journal} {ACS
  Synth. Biol.}\ }\textbf {\bibinfo {volume} {5}},\ \bibinfo {pages} {898}
  (\bibinfo {year} {2016})}\BibitemShut {NoStop}%
\bibitem [{\citenamefont {Fern}\ \emph {et~al.}(2017)\citenamefont {Fern},
  \citenamefont {Scalise}, \citenamefont {Cangialosi}, \citenamefont {Howie},
  \citenamefont {Potters},\ and\ \citenamefont {Schulman}}]{Fern2017-cd}%
  \BibitemOpen
  \bibfield  {author} {\bibinfo {author} {\bibfnamefont {J.}~\bibnamefont
  {Fern}}, \bibinfo {author} {\bibfnamefont {D.}~\bibnamefont {Scalise}},
  \bibinfo {author} {\bibfnamefont {A.}~\bibnamefont {Cangialosi}}, \bibinfo
  {author} {\bibfnamefont {D.}~\bibnamefont {Howie}}, \bibinfo {author}
  {\bibfnamefont {L.}~\bibnamefont {Potters}}, \ and\ \bibinfo {author}
  {\bibfnamefont {R.}~\bibnamefont {Schulman}},\ }\href {\doibase
  10.1021/acssynbio.6b00170} {\bibfield  {journal} {\bibinfo  {journal} {ACS
  Synth. Biol.}\ }\textbf {\bibinfo {volume} {6}},\ \bibinfo {pages} {190}
  (\bibinfo {year} {2017})}\BibitemShut {NoStop}%
\bibitem [{\citenamefont {Yordanov}\ \emph {et~al.}(2014)\citenamefont
  {Yordanov}, \citenamefont {Kim}, \citenamefont {Petersen}, \citenamefont
  {Shudy}, \citenamefont {Kulkarni},\ and\ \citenamefont
  {Phillips}}]{Yordanov2014-tn}%
  \BibitemOpen
  \bibfield  {author} {\bibinfo {author} {\bibfnamefont {B.}~\bibnamefont
  {Yordanov}}, \bibinfo {author} {\bibfnamefont {J.}~\bibnamefont {Kim}},
  \bibinfo {author} {\bibfnamefont {R.~L.}\ \bibnamefont {Petersen}}, \bibinfo
  {author} {\bibfnamefont {A.}~\bibnamefont {Shudy}}, \bibinfo {author}
  {\bibfnamefont {V.~V.}\ \bibnamefont {Kulkarni}}, \ and\ \bibinfo {author}
  {\bibfnamefont {A.}~\bibnamefont {Phillips}},\ }\href {\doibase
  10.1021/sb400169s} {\bibfield  {journal} {\bibinfo  {journal} {ACS Synth.
  Biol.}\ }\textbf {\bibinfo {volume} {3}},\ \bibinfo {pages} {600} (\bibinfo
  {year} {2014})}\BibitemShut {NoStop}%
\bibitem [{\citenamefont {Agrawal}\ \emph {et~al.}(2015)\citenamefont
  {Agrawal}, \citenamefont {Franco},\ and\ \citenamefont
  {Schulman}}]{Agrawal2015-bt}%
  \BibitemOpen
  \bibfield  {author} {\bibinfo {author} {\bibfnamefont {D.~K.}\ \bibnamefont
  {Agrawal}}, \bibinfo {author} {\bibfnamefont {E.}~\bibnamefont {Franco}}, \
  and\ \bibinfo {author} {\bibfnamefont {R.}~\bibnamefont {Schulman}},\ }\href
  {\doibase 10.1098/rsif.2015.0586} {\bibfield  {journal} {\bibinfo  {journal}
  {J. R. Soc. Interface}\ }\textbf {\bibinfo {volume} {12}},\ \bibinfo {pages}
  {20150586} (\bibinfo {year} {2015})}\BibitemShut {NoStop}%
\bibitem [{\citenamefont {Genot}\ \emph {et~al.}(2013)\citenamefont {Genot},
  \citenamefont {Bath},\ and\ \citenamefont {Turberfield}}]{Genot2013-rd}%
  \BibitemOpen
  \bibfield  {author} {\bibinfo {author} {\bibfnamefont {A.~J.}\ \bibnamefont
  {Genot}}, \bibinfo {author} {\bibfnamefont {J.}~\bibnamefont {Bath}}, \ and\
  \bibinfo {author} {\bibfnamefont {A.~J.}\ \bibnamefont {Turberfield}},\
  }\href {\doibase 10.1002/anie.201206201} {\bibfield  {journal} {\bibinfo
  {journal} {Angew. Chem. Int. Ed Engl.}\ }\textbf {\bibinfo {volume} {52}},\
  \bibinfo {pages} {1189} (\bibinfo {year} {2013})}\BibitemShut {NoStop}%
\bibitem [{\citenamefont {Virinchi}\ \emph {et~al.}(2017)\citenamefont
  {Virinchi}, \citenamefont {Behera},\ and\ \citenamefont
  {Gopalkrishnan}}]{Virinchi2017-pv}%
  \BibitemOpen
  \bibfield  {author} {\bibinfo {author} {\bibfnamefont {M.~V.}\ \bibnamefont
  {Virinchi}}, \bibinfo {author} {\bibfnamefont {A.}~\bibnamefont {Behera}}, \
  and\ \bibinfo {author} {\bibfnamefont {M.}~\bibnamefont {Gopalkrishnan}},\
  }in\ \href {\doibase 10.1007/978-3-319-66799-7\_6} {\emph {\bibinfo
  {booktitle} {{DNA} Computing and Molecular Programming}}}\ (\bibinfo
  {publisher} {Springer International Publishing},\ \bibinfo {year} {2017})\
  pp.\ \bibinfo {pages} {82--97}\BibitemShut {NoStop}%
\bibitem [{\citenamefont {Virinchi}\ \emph {et~al.}(2018)\citenamefont
  {Virinchi}, \citenamefont {Behera},\ and\ \citenamefont
  {Gopalkrishnan}}]{Virinchi2018-bw}%
  \BibitemOpen
  \bibfield  {author} {\bibinfo {author} {\bibfnamefont {M.~V.}\ \bibnamefont
  {Virinchi}}, \bibinfo {author} {\bibfnamefont {A.}~\bibnamefont {Behera}}, \
  and\ \bibinfo {author} {\bibfnamefont {M.}~\bibnamefont {Gopalkrishnan}},\
  }\href@noop {} {\  (\bibinfo {year} {2018})},\ \Eprint
  {http://arxiv.org/abs/1804.09062} {arXiv:1804.09062 [cs.ET]} \BibitemShut
  {NoStop}%
\bibitem [{\citenamefont {Zadorin}\ \emph {et~al.}(2015)\citenamefont
  {Zadorin}, \citenamefont {Rondelez}, \citenamefont {Galas},\ and\
  \citenamefont {Estevez-Torres}}]{ZadorinEstevez2015}%
  \BibitemOpen
  \bibfield  {author} {\bibinfo {author} {\bibfnamefont {A.~S.}\ \bibnamefont
  {Zadorin}}, \bibinfo {author} {\bibfnamefont {Y.}~\bibnamefont {Rondelez}},
  \bibinfo {author} {\bibfnamefont {J.-C.}\ \bibnamefont {Galas}}, \ and\
  \bibinfo {author} {\bibfnamefont {A.}~\bibnamefont {Estevez-Torres}},\ }\href
  {\doibase 10.1103/PhysRevLett.114.068301} {\bibfield  {journal} {\bibinfo
  {journal} {Phys. Rev. Lett.}\ }\textbf {\bibinfo {volume} {114}},\ \bibinfo
  {pages} {068301} (\bibinfo {year} {2015})}\BibitemShut {NoStop}%
\bibitem [{\citenamefont {Sarpeshkar}(2014)}]{Sarpeshkar2014-qx}%
  \BibitemOpen
  \bibfield  {author} {\bibinfo {author} {\bibfnamefont {R.}~\bibnamefont
  {Sarpeshkar}},\ }\href {\doibase 10.1098/rsta.2013.0110} {\bibfield
  {journal} {\bibinfo  {journal} {Philos. Trans. A Math. Phys. Eng. Sci.}\
  }\textbf {\bibinfo {volume} {372}},\ \bibinfo {pages} {20130110} (\bibinfo
  {year} {2014})}\BibitemShut {NoStop}%
\bibitem [{\citenamefont {Sarpeshkar}(1998)}]{Sarpeshkar1998-yx}%
  \BibitemOpen
  \bibfield  {author} {\bibinfo {author} {\bibfnamefont {R.}~\bibnamefont
  {Sarpeshkar}},\ }\href@noop {} {\bibfield  {journal} {\bibinfo  {journal}
  {Neural Comput.}\ }\textbf {\bibinfo {volume} {10}},\ \bibinfo {pages} {1601}
  (\bibinfo {year} {1998})}\BibitemShut {NoStop}%
\bibitem [{\citenamefont {Doty}(2013)}]{Doty2013}%
  \BibitemOpen
  \bibfield  {author} {\bibinfo {author} {\bibfnamefont {D.}~\bibnamefont
  {Doty}},\ }\href {http://arxiv.org/abs/1304.0872} {\bibfield  {journal}
  {\bibinfo  {journal} {CoRR}\ }\textbf {\bibinfo {volume} {abs/1304.0872}}
  (\bibinfo {year} {2013})},\ \Eprint {http://arxiv.org/abs/1304.0872}
  {arXiv:1304.0872} \BibitemShut {NoStop}%
\bibitem [{\citenamefont {Seelig}\ \emph {et~al.}(2006)\citenamefont {Seelig},
  \citenamefont {Soloveichik}, \citenamefont {Zhang},\ and\ \citenamefont
  {Winfree}}]{Seelig2006-tp}%
  \BibitemOpen
  \bibfield  {author} {\bibinfo {author} {\bibfnamefont {G.}~\bibnamefont
  {Seelig}}, \bibinfo {author} {\bibfnamefont {D.}~\bibnamefont {Soloveichik}},
  \bibinfo {author} {\bibfnamefont {D.~Y.}\ \bibnamefont {Zhang}}, \ and\
  \bibinfo {author} {\bibfnamefont {E.}~\bibnamefont {Winfree}},\ }\href
  {\doibase 10.1126/science.1132493} {\bibfield  {journal} {\bibinfo  {journal}
  {Science}\ }\textbf {\bibinfo {volume} {314}},\ \bibinfo {pages} {1585}
  (\bibinfo {year} {2006})}\BibitemShut {NoStop}%
\bibitem [{\citenamefont {Kishi}\ \emph {et~al.}(2018)\citenamefont {Kishi},
  \citenamefont {Schaus}, \citenamefont {Gopalkrishnan}, \citenamefont {Xuan},\
  and\ \citenamefont {Yin}}]{Kishi2018-qq}%
  \BibitemOpen
  \bibfield  {author} {\bibinfo {author} {\bibfnamefont {J.~Y.}\ \bibnamefont
  {Kishi}}, \bibinfo {author} {\bibfnamefont {T.~E.}\ \bibnamefont {Schaus}},
  \bibinfo {author} {\bibfnamefont {N.}~\bibnamefont {Gopalkrishnan}}, \bibinfo
  {author} {\bibfnamefont {F.}~\bibnamefont {Xuan}}, \ and\ \bibinfo {author}
  {\bibfnamefont {P.}~\bibnamefont {Yin}},\ }\href {\doibase
  10.1038/nchem.2872} {\bibfield  {journal} {\bibinfo  {journal} {Nat. Chem.}\
  }\textbf {\bibinfo {volume} {10}},\ \bibinfo {pages} {155} (\bibinfo {year}
  {2018})}\BibitemShut {NoStop}%
\bibitem [{\citenamefont {Alon}(2007)}]{Alon2007-qt}%
  \BibitemOpen
  \bibfield  {author} {\bibinfo {author} {\bibfnamefont {U.}~\bibnamefont
  {Alon}},\ }\href@noop {} {\bibfield  {journal} {\bibinfo  {journal} {Nat.
  Rev. Genet.}\ }\textbf {\bibinfo {volume} {8}},\ \bibinfo {pages} {450}
  (\bibinfo {year} {2007})}\BibitemShut {NoStop}%
\bibitem [{\citenamefont {Cuba~Samaniego}\ and\ \citenamefont
  {Franco}(2018)}]{Cuba_Samaniego2018-ja}%
  \BibitemOpen
  \bibfield  {author} {\bibinfo {author} {\bibfnamefont {C.}~\bibnamefont
  {Cuba~Samaniego}}\ and\ \bibinfo {author} {\bibfnamefont {E.}~\bibnamefont
  {Franco}},\ }\href {\doibase 10.1021/acssynbio.7b00222} {\bibfield  {journal}
  {\bibinfo  {journal} {ACS Synth. Biol.}\ }\textbf {\bibinfo {volume} {7}},\
  \bibinfo {pages} {75} (\bibinfo {year} {2018})}\BibitemShut {NoStop}%
\bibitem [{\citenamefont {Graves}\ \emph {et~al.}(2013)\citenamefont {Graves},
  \citenamefont {r.~Mohamed},\ and\ \citenamefont {Hinton}}]{Graves2013-qs}%
  \BibitemOpen
  \bibfield  {author} {\bibinfo {author} {\bibfnamefont {A.}~\bibnamefont
  {Graves}}, \bibinfo {author} {\bibfnamefont {A.}~\bibnamefont {r.~Mohamed}},
  \ and\ \bibinfo {author} {\bibfnamefont {G.}~\bibnamefont {Hinton}},\ }in\
  \href {\doibase 10.1109/ICASSP.2013.6638947} {\emph {\bibinfo {booktitle}
  {2013 {IEEE} International Conference on Acoustics, Speech and Signal
  Processing}}}\ (\bibinfo {year} {2013})\ pp.\ \bibinfo {pages}
  {6645--6649}\BibitemShut {NoStop}%
\bibitem [{\citenamefont {Tu}(2013)}]{Tu2013-sk}%
  \BibitemOpen
  \bibfield  {author} {\bibinfo {author} {\bibfnamefont {Y.}~\bibnamefont
  {Tu}},\ }\href@noop {} {\bibfield  {journal} {\bibinfo  {journal} {Annu. Rev.
  Biophys.}\ }\textbf {\bibinfo {volume} {42}},\ \bibinfo {pages} {337}
  (\bibinfo {year} {2013})}\BibitemShut {NoStop}%
\bibitem [{\citenamefont {Ferrell}(2016)}]{FERRELL201662}%
  \BibitemOpen
  \bibfield  {author} {\bibinfo {author} {\bibfnamefont {J.~E.}\ \bibnamefont
  {Ferrell}},\ }\href {\doibase https://doi.org/10.1016/j.cels.2016.02.006}
  {\bibfield  {journal} {\bibinfo  {journal} {Cell Systems}\ }\textbf {\bibinfo
  {volume} {2}},\ \bibinfo {pages} {62 } (\bibinfo {year} {2016})}\BibitemShut
  {NoStop}%
\bibitem [{\citenamefont {Chirieleison}\ \emph {et~al.}(2013)\citenamefont
  {Chirieleison}, \citenamefont {Allen}, \citenamefont {Simpson}, \citenamefont
  {Ellington},\ and\ \citenamefont {Chen}}]{Chirieleison_2013}%
  \BibitemOpen
  \bibfield  {author} {\bibinfo {author} {\bibfnamefont {S.~M.}\ \bibnamefont
  {Chirieleison}}, \bibinfo {author} {\bibfnamefont {P.~B.}\ \bibnamefont
  {Allen}}, \bibinfo {author} {\bibfnamefont {Z.~B.}\ \bibnamefont {Simpson}},
  \bibinfo {author} {\bibfnamefont {A.~D.}\ \bibnamefont {Ellington}}, \ and\
  \bibinfo {author} {\bibfnamefont {X.}~\bibnamefont {Chen}},\ }\href {\doibase
  10.1038/nchem.1764} {\bibfield  {journal} {\bibinfo  {journal} {Nature
  Chemistry}\ }\textbf {\bibinfo {volume} {5}},\ \bibinfo {pages} {1000}
  (\bibinfo {year} {2013})}\BibitemShut {NoStop}%
\bibitem [{\citenamefont {Scalise}\ \emph {et~al.}(2018)\citenamefont
  {Scalise}, \citenamefont {Dutta},\ and\ \citenamefont
  {Schulman}}]{Scalise2018-jc}%
  \BibitemOpen
  \bibfield  {author} {\bibinfo {author} {\bibfnamefont {D.}~\bibnamefont
  {Scalise}}, \bibinfo {author} {\bibfnamefont {N.}~\bibnamefont {Dutta}}, \
  and\ \bibinfo {author} {\bibfnamefont {R.}~\bibnamefont {Schulman}},\ }\href
  {\doibase 10.1021/jacs.8b05373} {\bibfield  {journal} {\bibinfo  {journal}
  {J. Am. Chem. Soc.}\ }\textbf {\bibinfo {volume} {140}},\ \bibinfo {pages}
  {12069} (\bibinfo {year} {2018})}\BibitemShut {NoStop}%
\bibitem [{\citenamefont {Barkai}\ and\ \citenamefont
  {Leibler}(1997)}]{Barkai1997-zq}%
  \BibitemOpen
  \bibfield  {author} {\bibinfo {author} {\bibfnamefont {N.}~\bibnamefont
  {Barkai}}\ and\ \bibinfo {author} {\bibfnamefont {S.}~\bibnamefont
  {Leibler}},\ }\href {\doibase 10.1038/43199} {\bibfield  {journal} {\bibinfo
  {journal} {Nature}\ }\textbf {\bibinfo {volume} {387}},\ \bibinfo {pages}
  {913} (\bibinfo {year} {1997})}\BibitemShut {NoStop}%
\bibitem [{\citenamefont {Mitchell}\ \emph {et~al.}(2015)\citenamefont
  {Mitchell}, \citenamefont {Wei},\ and\ \citenamefont
  {Lim}}]{Mitchell2015-oa}%
  \BibitemOpen
  \bibfield  {author} {\bibinfo {author} {\bibfnamefont {A.}~\bibnamefont
  {Mitchell}}, \bibinfo {author} {\bibfnamefont {P.}~\bibnamefont {Wei}}, \
  and\ \bibinfo {author} {\bibfnamefont {W.~A.}\ \bibnamefont {Lim}},\ }\href
  {\doibase 10.1126/science.aab0892} {\bibfield  {journal} {\bibinfo  {journal}
  {Science}\ }\textbf {\bibinfo {volume} {350}},\ \bibinfo {pages} {1379}
  (\bibinfo {year} {2015})}\BibitemShut {NoStop}%
\bibitem [{\citenamefont {Sorre}\ \emph {et~al.}(2014)\citenamefont {Sorre},
  \citenamefont {Warmflash}, \citenamefont {Brivanlou},\ and\ \citenamefont
  {Siggia}}]{Sorre2014-oc}%
  \BibitemOpen
  \bibfield  {author} {\bibinfo {author} {\bibfnamefont {B.}~\bibnamefont
  {Sorre}}, \bibinfo {author} {\bibfnamefont {A.}~\bibnamefont {Warmflash}},
  \bibinfo {author} {\bibfnamefont {A.~H.}\ \bibnamefont {Brivanlou}}, \ and\
  \bibinfo {author} {\bibfnamefont {E.~D.}\ \bibnamefont {Siggia}},\ }\href
  {\doibase 10.1016/j.devcel.2014.05.022} {\bibfield  {journal} {\bibinfo
  {journal} {Dev. Cell}\ }\textbf {\bibinfo {volume} {30}},\ \bibinfo {pages}
  {334} (\bibinfo {year} {2014})}\BibitemShut {NoStop}%
\bibitem [{\citenamefont {Lulu~Qian}(2011)}]{Qian2011-Simple}%
  \BibitemOpen
  \bibfield  {author} {\bibinfo {author} {\bibfnamefont {E.~W.}\ \bibnamefont
  {Lulu~Qian}},\ }\href@noop {} {\bibfield  {journal} {\bibinfo  {journal}
  {Journal of The Royal Society Interface}\ }\textbf {\bibinfo {volume} {8}},\
  \bibinfo {pages} {1281} (\bibinfo {year} {2011})}\BibitemShut {NoStop}%
\bibitem [{\citenamefont {David~Soloveichik}(2010)}]{DNAUniversal}%
  \BibitemOpen
  \bibfield  {author} {\bibinfo {author} {\bibfnamefont {E.~W.}\ \bibnamefont
  {David~Soloveichik}, \bibfnamefont {Georg~Seelig}},\ }\href@noop {}
  {\bibfield  {journal} {\bibinfo  {journal} {Proceedings of the National
  Academy of Sciences}\ }\textbf {\bibinfo {volume} {107}},\ \bibinfo {pages}
  {5393} (\bibinfo {year} {2010})}\BibitemShut {NoStop}%
\bibitem [{\citenamefont {Soloveichik}\ \emph {et~al.}(2010)\citenamefont
  {Soloveichik}, \citenamefont {Seelig},\ and\ \citenamefont
  {Winfree}}]{Soloveichik2010-kz}%
  \BibitemOpen
  \bibfield  {author} {\bibinfo {author} {\bibfnamefont {D.}~\bibnamefont
  {Soloveichik}}, \bibinfo {author} {\bibfnamefont {G.}~\bibnamefont {Seelig}},
  \ and\ \bibinfo {author} {\bibfnamefont {E.}~\bibnamefont {Winfree}},\ }\href
  {\doibase 10.1073/pnas.0909380107} {\bibfield  {journal} {\bibinfo  {journal}
  {Proc. Natl. Acad. Sci. U. S. A.}\ }\textbf {\bibinfo {volume} {107}},\
  \bibinfo {pages} {5393} (\bibinfo {year} {2010})}\BibitemShut {NoStop}%
\bibitem [{\citenamefont {Zhang}\ and\ \citenamefont
  {Seelig}(2011)}]{Zhang2011-by}%
  \BibitemOpen
  \bibfield  {author} {\bibinfo {author} {\bibfnamefont {D.~Y.}\ \bibnamefont
  {Zhang}}\ and\ \bibinfo {author} {\bibfnamefont {G.}~\bibnamefont {Seelig}},\
  }\href@noop {} {\bibfield  {journal} {\bibinfo  {journal} {Nat. Chem.}\
  }\textbf {\bibinfo {volume} {3}},\ \bibinfo {pages} {103} (\bibinfo {year}
  {2011})}\BibitemShut {NoStop}%
\bibitem [{\citenamefont {Machinek}\ \emph {et~al.}(2014)\citenamefont
  {Machinek}, \citenamefont {Ouldridge}, \citenamefont {Haley}, \citenamefont
  {Bath},\ and\ \citenamefont {Turberfield}}]{Machinek2014-di}%
  \BibitemOpen
  \bibfield  {author} {\bibinfo {author} {\bibfnamefont {R.~R.~F.}\
  \bibnamefont {Machinek}}, \bibinfo {author} {\bibfnamefont {T.~E.}\
  \bibnamefont {Ouldridge}}, \bibinfo {author} {\bibfnamefont {N.~E.~C.}\
  \bibnamefont {Haley}}, \bibinfo {author} {\bibfnamefont {J.}~\bibnamefont
  {Bath}}, \ and\ \bibinfo {author} {\bibfnamefont {A.~J.}\ \bibnamefont
  {Turberfield}},\ }\href {\doibase 10.1038/ncomms6324} {\bibfield  {journal}
  {\bibinfo  {journal} {Nat. Commun.}\ }\textbf {\bibinfo {volume} {5}},\
  \bibinfo {pages} {5324} (\bibinfo {year} {2014})}\BibitemShut {NoStop}%
\bibitem [{\citenamefont {Ouldridge}(2015)}]{Ouldridge2015}%
  \BibitemOpen
  \bibfield  {author} {\bibinfo {author} {\bibfnamefont {T.~E.}\ \bibnamefont
  {Ouldridge}},\ }\href {\doibase 10.1080/00268976.2014.975293} {\bibfield
  {journal} {\bibinfo  {journal} {Molecular Physics}\ }\textbf {\bibinfo
  {volume} {113}},\ \bibinfo {pages} {1} (\bibinfo {year} {2015})},\ \Eprint
  {http://arxiv.org/abs/https://doi.org/10.1080/00268976.2014.975293}
  {https://doi.org/10.1080/00268976.2014.975293} \BibitemShut {NoStop}%
\bibitem [{\citenamefont {Benzinger}\ and\ \citenamefont
  {Khammash}(2018)}]{Benzinger2018-pf}%
  \BibitemOpen
  \bibfield  {author} {\bibinfo {author} {\bibfnamefont {D.}~\bibnamefont
  {Benzinger}}\ and\ \bibinfo {author} {\bibfnamefont {M.}~\bibnamefont
  {Khammash}},\ }\href {\doibase 10.1038/s41467-018-05882-2} {\bibfield
  {journal} {\bibinfo  {journal} {Nat. Commun.}\ }\textbf {\bibinfo {volume}
  {9}},\ \bibinfo {pages} {3521} (\bibinfo {year} {2018})}\BibitemShut
  {NoStop}%
\bibitem [{\citenamefont {Linko}\ \emph {et~al.}(2015)\citenamefont {Linko},
  \citenamefont {Ora},\ and\ \citenamefont {Kostiainen}}]{LINKO2015586}%
  \BibitemOpen
  \bibfield  {author} {\bibinfo {author} {\bibfnamefont {V.}~\bibnamefont
  {Linko}}, \bibinfo {author} {\bibfnamefont {A.}~\bibnamefont {Ora}}, \ and\
  \bibinfo {author} {\bibfnamefont {M.~A.}\ \bibnamefont {Kostiainen}},\ }\href
  {\doibase https://doi.org/10.1016/j.tibtech.2015.08.001} {\bibfield
  {journal} {\bibinfo  {journal} {Trends in Biotechnology}\ }\textbf {\bibinfo
  {volume} {33}},\ \bibinfo {pages} {586 } (\bibinfo {year}
  {2015})}\BibitemShut {NoStop}%
\bibitem [{\citenamefont {Dai}\ \emph {et~al.}(2012)\citenamefont {Dai},
  \citenamefont {Vorselen}, \citenamefont {Korolev},\ and\ \citenamefont
  {Gore}}]{Dai1175}%
  \BibitemOpen
  \bibfield  {author} {\bibinfo {author} {\bibfnamefont {L.}~\bibnamefont
  {Dai}}, \bibinfo {author} {\bibfnamefont {D.}~\bibnamefont {Vorselen}},
  \bibinfo {author} {\bibfnamefont {K.~S.}\ \bibnamefont {Korolev}}, \ and\
  \bibinfo {author} {\bibfnamefont {J.}~\bibnamefont {Gore}},\ }\href {\doibase
  10.1126/science.1219805} {\bibfield  {journal} {\bibinfo  {journal}
  {Science}\ }\textbf {\bibinfo {volume} {336}},\ \bibinfo {pages} {1175}
  (\bibinfo {year} {2012})},\ \Eprint
  {http://arxiv.org/abs/http://science.sciencemag.org/content/336/6085/1175.full.pdf}
  {http://science.sciencemag.org/content/336/6085/1175.full.pdf} \BibitemShut
  {NoStop}%
\bibitem [{\citenamefont {Lakin}\ and\ \citenamefont
  {Stefanovic}(2016)}]{Lakin2016-jv}%
  \BibitemOpen
  \bibfield  {author} {\bibinfo {author} {\bibfnamefont {M.~R.}\ \bibnamefont
  {Lakin}}\ and\ \bibinfo {author} {\bibfnamefont {D.}~\bibnamefont
  {Stefanovic}},\ }\href {\doibase 10.1021/acssynbio.6b00009} {\bibfield
  {journal} {\bibinfo  {journal} {ACS Synth. Biol.}\ }\textbf {\bibinfo
  {volume} {5}},\ \bibinfo {pages} {885} (\bibinfo {year} {2016})}\BibitemShut
  {NoStop}%
\bibitem [{\citenamefont {Lakin}\ \emph {et~al.}(2014)\citenamefont {Lakin},
  \citenamefont {Minnich}, \citenamefont {Lane},\ and\ \citenamefont
  {Stefanovic}}]{Lakin2014-qp}%
  \BibitemOpen
  \bibfield  {author} {\bibinfo {author} {\bibfnamefont {M.~R.}\ \bibnamefont
  {Lakin}}, \bibinfo {author} {\bibfnamefont {A.}~\bibnamefont {Minnich}},
  \bibinfo {author} {\bibfnamefont {T.}~\bibnamefont {Lane}}, \ and\ \bibinfo
  {author} {\bibfnamefont {D.}~\bibnamefont {Stefanovic}},\ }\href@noop {}
  {\bibfield  {journal} {\bibinfo  {journal} {J. R. Soc. Interface}\ }\textbf
  {\bibinfo {volume} {11}},\ \bibinfo {pages} {20140902} (\bibinfo {year}
  {2014})}\BibitemShut {NoStop}%
\bibitem [{\citenamefont {Poole}\ \emph {et~al.}(2017)\citenamefont {Poole},
  \citenamefont {Ortiz-Mu{\~n}oz}, \citenamefont {Behera}, \citenamefont
  {Jones}, \citenamefont {Ouldridge}, \citenamefont {Winfree},\ and\
  \citenamefont {Gopalkrishnan}}]{Poole2017-cz}%
  \BibitemOpen
  \bibfield  {author} {\bibinfo {author} {\bibfnamefont {W.}~\bibnamefont
  {Poole}}, \bibinfo {author} {\bibfnamefont {A.}~\bibnamefont
  {Ortiz-Mu{\~n}oz}}, \bibinfo {author} {\bibfnamefont {A.}~\bibnamefont
  {Behera}}, \bibinfo {author} {\bibfnamefont {N.~S.}\ \bibnamefont {Jones}},
  \bibinfo {author} {\bibfnamefont {T.~E.}\ \bibnamefont {Ouldridge}}, \bibinfo
  {author} {\bibfnamefont {E.}~\bibnamefont {Winfree}}, \ and\ \bibinfo
  {author} {\bibfnamefont {M.}~\bibnamefont {Gopalkrishnan}},\ }in\ \href
  {\doibase 10.1007/978-3-319-66799-7\_14} {\emph {\bibinfo {booktitle} {{DNA}
  Computing and Molecular Programming}}}\ (\bibinfo  {publisher} {Springer
  International Publishing},\ \bibinfo {year} {2017})\ pp.\ \bibinfo {pages}
  {210--231}\BibitemShut {NoStop}%
\bibitem [{\citenamefont {Zhang}\ and\ \citenamefont
  {Winfree}(2009)}]{ControlofDNAWinfree}%
  \BibitemOpen
  \bibfield  {author} {\bibinfo {author} {\bibfnamefont {D.~Y.}\ \bibnamefont
  {Zhang}}\ and\ \bibinfo {author} {\bibfnamefont {E.}~\bibnamefont
  {Winfree}},\ }\href {\doibase 10.1021/ja906987s} {\bibfield  {journal}
  {\bibinfo  {journal} {Journal of the American Chemical Society}\ }\textbf
  {\bibinfo {volume} {131}},\ \bibinfo {pages} {17303} (\bibinfo {year}
  {2009})},\ \bibinfo {note} {pMID: 19894722},\ \Eprint
  {http://arxiv.org/abs/https://doi.org/10.1021/ja906987s}
  {https://doi.org/10.1021/ja906987s} \BibitemShut {NoStop}%
\end{thebibliography}%

\end{document}